\documentclass[%
 reprint,
 amsmath,amssymb,
 aip,
]{revtex4-2}

\usepackage{graphicx}
\usepackage{dcolumn}
\usepackage{bm}

\begin{document}


\title{Reactor scale simulations of ALD and ALE: ideal and non-ideal self-limited processes in a cylindrical and a 300 mm wafer cross-flow reactor}

\author{Angel Yanguas-Gil}
\email{ayg@anl.gov}

\author{Joseph A. Libera}

\author{Jeffrey W. Elam}%
\affiliation{%
 Applied Materials Division\\
 Argonne National Laboratory, IL 60439 (USA)
}%

\date{\today}

\begin{abstract}
We have developed a simulation tool to model
self-limited processes such as atomic layer
deposition and atomic layer etching inside reactors
of arbitrary geometry. In this work, we have
applied this model to two standard types of cross-flow
reactors: a cylindrical reactor and a model 300 mm wafer
reactor, and explored both ideal and non-ideal 
self-limited kinetics. For the cylindrical tube reactor the full simulation 
results agree well with analytic expressions obtained using 
a simple plug flow model, though the presence
of axial diffusion tends to soften growth profiles with respect to the plug
flow case. Our simulations also allowed us to model the output of 
in-situ techniques such as quartz crystal microbalance and
mass spectrometry, providing a way of discriminating between
ideal and non-ideal surface kinetics using in-situ measurements.  We extended the simulations
to consider two non-ideal self-limited processes: soft-saturating processes characterized by
a slow reaction pathway, and processes where surface byproducts can compete with the precursor
for the same pool of adsorption sites, allowing us to
quantify their impact 
in the thickness variability across 300 mm wafer substrates.

\end{abstract}

\maketitle


\section{\label{sec:intro}Introduction}

Atomic layer deposition (ALD) is a thin film growth technique
that has been experiencing a tremendous growth in
terms of processes and application domains, ranging
from semiconductor manufacturing to photovoltaics
and energy storage. More recently, atomic layer
etching (ALE), ALD's etching counterpart, has
also seen a resurgence due to its impact for 
advanced lithographic applications in semiconductor
processing and the development of thermal ALE processes.\cite{George_ALE_2016}
ALD and ALE are both enabled by self-limited precursor-surface
interactions, which underpin their ability to
add/remove material in a controlled way over large 
substrate areas and inside nanostructured and high 
aspect ratio features.

ALD and ALE's self-limited nature is usually conceptualized in
terms of a finite number of reactive sites on the
surface: once the reactive sites are fully consumed,
the surface is no longer reactive towards the precursor. However,
this picture overlooks some critical aspects of self-limited
processes: first, self-limited heterogeneous reactions introduce
 non-linear time-dependent kinetics and complex spatio-temporal patterns in the 
surface reactivity and chemistry during each precursor
dose. Consequently, the contribution of 
the reactive transport of species inside the reactor  needs
to be considered when extracting information
on the surface kinetics from in-situ measurements. The transport
of species is also key to understanding the scale-up
of a process and optimizing reactor geometry.

Second, the self-limited nature of precursor-surface interactions
is not sufficient to guarantee uniformity over large areas or
inside high aspect ratio features: even without considering
the effect of additional non self-limiting components, the
presence of soft-
saturating kinetics, additional surface
recombination pathways, competitive adsorption with 
reaction byproducts, flux/pressure dependent surface kinetics,
long surface residence times, or merely insufficient purge times,
can lead to inhomogeneous processes even when the processes
are self-limited. As 
mentioned in a recent review,\cite{Sonsteby2020} these processes can also affect
the reproducibility of ALD and, by extension, ALE.

In this work, we explore the reactive  transport of species
under self-limited processes using computational fluid
dynamics (CFD) simulations. CFD is a valuable tool 
to predict the impact of surface kinetics and reactor
geometry has on metrics such as
throughput, coating homogeneity, and precursor utilization.
Examples in the literature have 
explored CFD and multiscale simulations to model and
optimize ALD processes.\cite{Sengupta_multiscale_2005,Holmqvist_models_2012,Gakis_simulations_2018,Peltonen_simulations_2018}
Here we use CFD to
address the following
three questions: 1) how can we
use under-saturated doses to extract
information on surface kinetics from coverage and thickness profiles?
2) how does reactive
transport impact the data obtained using in-situ measurement tools
such as quartz crystal microbalance (QCM)
and quadrupole mass spectrometry (QMS), and 3) what
is the impact that non-idealities in the surface
kinetics have on the saturation profiles and homogeneity
at the reactor scale?
Understanding these three aspects is crucial to maximize
the information that we 
can extract from reactor-scale data.

Finally, one of our goals is to make the simulation code available to the research community: the simulation tools and some
examples that can be used to reproduce the results presented in this manuscript have been publicly released as open source in github (https://github.com/aldsim/aldFoam) under a GPLv3 license.

\section{\label{sec:model}Model}

\subsection{Model equations}

Our model solves the time-dependent 
reactive transport of one or more reactive species subject
to self-limited heterogeneous processes inside
reactors of arbitrary geometry.
Our
model makes two fundamental assumptions: 1) we
assume that precursors and reaction byproducts represent
a small perturbation with respect
to the carrier gas flow and 2) we consider that the total carrier gas flow is minimally altered during doses and purge times.  Both
approximations are consistent with the way our experimental reactors operate: the carrier gas lines utilized during the precursor
dose and purge operations in 
our reactor have similar conductances,
so the effect of bypassing a precursor
bubbler during the purge times on the overall flow is minimal.\cite{Elam_viscousflow_2002}

These two assumptions
allow us to decouple the momentum and energy transport from the precursor mass transport equation, so that the reactive
transport of both precursors and reactants takes place on a velocity field $\bm{u}(\bm{x})$ that
is determined by the carrier gas and the overall process conditions.
Under these assumptions, we can
approximate the momentum transport equation with the incompressible
Navier-Stokes equation for the carrier gas:
 \begin{equation}
 \frac{\partial \bm{u}}{\partial  t } + (\bm{u} \cdot \nabla)\bm{u} -
 \nu \nabla^2 \bm{u} = -\frac{1}{\rho} \nabla p
\end{equation}
 Where $\nu = \mu/\rho$ is the kinematic 
 viscosity of the carrier gas, $\mu$ is the dynamic
 viscosity, and $\rho$ is the mass density.
Note that this
approximation assumes an isothermal reactor, since the
kinematic viscosity depends on the temperature both
through the dynamic viscosity and gas density. 
This is a restriction 
that can be easily lifted if the temperature dependence with
the surface kinetic parameters is known.
A discussion on non-isothermal conditions is given in Section \ref{sec_implementation}.

Under steady-state conditions, the solution of the Navier-Stokes
equations results in a velocity field $\bm{u} = \bm{u}(\bm{x})$ that is then used
as input for the mass transport equations of each of the chemical
vapor species.
The mass transport equation for each species
can be formulated in terms of
its molar concentration $c_i$ as:
\begin{equation}
\label{eq_ci}
\frac{\partial c_i}{\partial  t } + \nabla \cdot (\mathbf{u} c_i) = -\nabla \cdot \mathbf{J}_i
\end{equation}
with
\begin{equation}
\mathbf{J}_i = - D_i \nabla c_i
\end{equation}
Here $D_i$ is the diffusivity of species $i$ in the carrier gas.

The molar concentration $c_i$ can be expressed in terms of the precursor pressure $p_i$ as:
\begin{equation}
    \label{eq_cipi}
    c_i = \frac{p_i}{RT}
\end{equation}
where $R$ is the gas constant.

Our model neglects gas phase reactions. However,
the presence of non self-limited processes can be directly incorporated through the boundary conditions. They
also spontaneously
occur whenever the precursor and co-reactant are
simultaneously present in the gas phase.

\subsection{Boundary conditions}

We codify the reactivity of each species $i$ in terms of a wall reaction probability $\beta_i$. This terms incorporate reversible and
irreversible interactions as well as any side reaction pathways that do not contribute to either growth or etching, such as
surface recombination.  The second term that we need to consider is
desorption from the walls. We 
codify these processes in terms of a flux $F_i$, so that the mass balance equation at each point of the reactor surface can
be expressed as:
\begin{equation}
- D \frac{\partial c_i}{\partial n} = \beta_i \frac{1}{4}v_{th,i} c_i - F_i
\end{equation}

The exact dependence of $\beta_i$ and $F_i$ on the surface kinetics will vary depending on the specific heterogeneous processes being
considered in the model. These will be described in more detail in Section \ref{sec:reactions}. Also note that, with this formalism, the reactive transport
equations are the same regardless of the type of process that we are modeling, encompassing both self-limited growth (ALD) and etching (ALE).

Inlet boundary conditions need to incorporate the pulsed nature of ALD and ALE. A key challenge is how to model accurate pulses when the
the reactor inlet is not considered in the simulation domain. Here we have assumed concentration pulses at the inlet that are characterized by a
response time $t_r$ that controls the steepness of the pulse:
at the beginning of each
dose, the precursor concentration increases linearly during a time interval $t_r$ 
and remains fixed at a preset value $c_0$ during a time equal to $t_d-t_r$. Then
the concentration decreases linearly with time over the same internal $t_r$. This
allows us to control the spread of the pulse at the inlet while keeping the
product $t_d \times p_0$ constant, where $p_0$ refers to the precursor pressure
at the inlet. For reaction byproducts ($c_{bp}$), we assume that there
is no net mass transport
at the inlet, by imposing the condition:
\begin{equation}
    \bm{u}|_n c_{bp}  = - D_{bp}\frac{\partial c_0}{\partial n}
\end{equation}
Zero gradient boundary conditions were used for the outlet.

\subsection{\label{sec:reactions} Surface models for self-limited processes}

\subsubsection{\label{sec_ideal} Ideal self-limited process}

A key assumption of self-limited processes is 
the presence of a finite number of surface sites. If we define the surface fractional
coverage $\Theta$ as the fraction of surface sites that have reacted with the
precursor,
the simplest model is to assume that the reactivity of the precursor is given by:\cite{YanguasGil2014,yanguasgil2012a}
\begin{equation}
    \label{eq_betaideal}
    \beta_1 = \beta_{10}(1-\Theta)
\end{equation}
that is, the surface reactivity towards the precursor is proportional to the fraction of available sites. This model corresponds to an irreversible first order Langmuir
kinetics, and it is one of the most widely used models for ALD simulations. The
evolution with time of the surface coverage will be then given by:
\begin{equation}
    \label{eq_ideal}
    \frac{d\Theta}{dt} = s_0 \beta_{10} J_1 (1-\Theta)
\end{equation}

When we consider full ALD cycles with both precursor and co-reactant doses,
the evolution of the fractional coverage simply incorporates the influence
of both species:
\begin{equation}
    \label{eq_ideal2}
    \frac{d\Theta}{dt} = s_0 n_2 \beta_{10} J_1 (1-\Theta)
    - s_0\beta_{20} J_2 \Theta
\end{equation}
Here $J_1$ and $J_2$ are the surface flux of precursor species to the surface, and $s_0$ is the average surface area of a reaction site, and $n_2$ is
the number of co-reactant molecules required per precursor molecule
required to satisfy the film's stoichiometry.

The wall flux $J_i$ can be expressed as a function of the precursor pressure as:
\begin{equation}
J_i = \frac{1}{4}v_{th}\frac{p_i}{kT}
\end{equation} 
where $v_{th}$ is the mean thermal velocity of
species $i$
and $p_i$ is connected with the molar density $c_i$
through Eq. \ref{eq_cipi}.

The value of $s_0$ can be obtained from the
mass gain 
(or loss in the case of etching) per unit surface area per
cycle $\Delta m$ or the growth (etch) per cycle in
unit of thickness. For the former, $s_0$ is simply given by:
\begin{equation}
    s_0 = \frac{M_0}{n_p \Delta m} 
\end{equation}
where $M_0$ is the molecular mass of the solid, and
$n_p$ is the number of precursor molecules per
unit formula (i.e. 2 in the case of trimethylaluminum and Al$_2$O$_3$).

\subsubsection{\label{sec_softsat} Soft-saturating processes}

A simple generalization to the ideal model is
to consider two self-limited reaction pathways with
a fast and a slow reacting component.\cite{YanguasGil2014b}
This allows us to incorporate soft-saturating processes where saturation is not reached as fast as in the ideal case.

If $f$ represents the fraction of sites with the second pathway, the reaction probability 
$\beta_1$ will be given by:
\begin{equation}
    \beta_1 = (1-f) \beta_{1a}(1-\Theta_a) + f\beta_{1b}(1-\Theta_b)
\end{equation}
where $\beta_{1a}$ and $\beta_{1b}$ represent
the sticking probabilities for each pathway and
$\Theta_a$ and $\Theta_b$ their respective fractional
coverages, so that the total surface coverage will be:
\begin{equation}
    \Theta = (1-f)\Theta_a + f \Theta_b
\end{equation}
and the evolution of $\Theta_a$ and $\Theta_b$ is
given by Eq. \ref{eq_ideal}.

\subsubsection{\label{sec_site} Site-blocking by precursor byproducts}

Thus far we have assumed that heterogeneous
processes involve a precursor and a
co-reactant species. However, reaction
byproducts and precursor ligands
can play an important role, for instance 
competing with precursor molecules for available
surface sites. In the case of ALD, this can
lead to the presence of thickness gradients in
the reactor even under saturation with otherwise perfectly self-limited processes.\cite{Ritala_TTIP_1993,Elers_uniformityreview_2006}

Here we have implemented a simple model that captures the impact of precursor 
byproducts through a simple
site-blocking mechanism. Our model considers the surface fractional coverage of both the precursor
and precursor byproducts, $\Theta$ and $\Theta_{bp}$,
respectively.

The simplest possible model assumes that upon adsorption, the precursor occupies a single
surface site, releasing $n_{bp}$ reaction byproducts into
the gas phase. These byproducts can then adsorb
on available sites. The co-reactant is then
able of fully regenerating the surface. This
can be modeled using the following equations:
\begin{equation}
    \frac{d\Theta}{dt} = s_0 \beta_{10} J_1 (1-\Theta-\Theta_{bp})
\end{equation}
\begin{equation}
    \frac{d\Theta_{bp}}{dt} = s_0 \beta_{bp0} J_{bp} (1-\Theta-\Theta_{bp})
\end{equation}
In the case of one dose. The flux of byproduct molecules coming back to the
gas phase is given by:
\begin{equation}
F_{bp} = n_{bp} s_0 \beta_{10} J_1 (1-\Theta-\Theta_{bp})
\end{equation}

When two doses are considered, we have to further consider the 
\begin{equation}
    \frac{d\Theta}{dt} = s_0 \beta_{10} J_1 (1-\Theta-\Theta_{bp})-s_0 n_2 \beta_{20} J_2 \Theta
\end{equation}
\begin{equation}
    \frac{d\Theta_{bp}}{dt} = s_0 \beta_{bp0} J_{bp} (1-\Theta-\Theta_{bp})-s_0 \beta_{20} J_2 \Theta_{bp}
\end{equation}
In the case of simulations considering full cycles.

\subsubsection{Non self-limited and recombination pathways}

Another way of generalizing Eq. \ref{eq_ideal} is to
 consider non self-limited as well as secondary
 pathways such as surface recombination that
 do not lead to either growth or etching.
We can implement them simply by adding
extra components to the reaction probability:
\begin{equation}
    \beta_1 = \beta_{10}(1-\Theta) + \beta_2 + \beta_{rec}
\end{equation}

\subsubsection{Sticky precursors}

The final case that we can consider are precursors
that have a significant residence time on the
surface. These may require larger
purge times in order to fully evacuate unreacted
precursor molecules from the reactor.

Here we consider a simple model, where a precursor
molecule can undergo monolayer adsorption on reacted
sites. Under this assumption, we have available
sites, reacted sites that don't have adsorbed
precursor molecules, whose fractional coverage
is given by $\Theta$, and reacted sites with
adsorbed precursor molecules, whose fractional
coverage is given by $\Theta_a$. The evolution
of $\Theta$ and $\Theta_a$ will be given by:
\begin{equation}
    \frac{d\Theta}{dt} = s_0 \beta_{10} J_1 (1-\Theta-\Theta_a) + k_0 \Theta_a
\end{equation}
\begin{equation}
    \frac{d\Theta_a}{dt} = s_0 \beta_{a0} J_a \Theta - k_d \Theta_a
\end{equation}
here $k_d$ is the desorption rate, and $\beta_{a0}$
is the sticking probability for reversible
adsorption.

A consequence of having this process is that the growth
per cycle becomes larger than the nominal saturation value when
the coreactant reacts with reversibly adsorbed precursor
molecules.

\subsection{\label{sec_implementation} Implementation}

We implemented the models described above
using the open source library OpenFOAM.\cite{Weller_OpenFOAM_1998}
OpenFOAM solves partial differential equations using a finite volume method
with a co-located grid
approach in which all properties are stored at a single point of each control volume (its centroid). Interpolation,
discretization, and matrix solution schemes can be selected at runtime. Through OpenFOAM we
 can work with arbitrary reactor geometries, including 1D, 2D, 2D axisymmetric, and full 3D simulations.

In order to incorporate self-limited processes, we have created a series of
custom solvers for both ideal and non-ideal self-limited interactions. Volume
fields such as reactant and byproduct concentrations are still solved and
discretized using OpenFOAM's built-in capabilities. The time evolution of the different surface coverages are solved using
a custom solver. Each
of the active boundaries is initiated using kinetic parameters that are specific to each region in our mesh. This allows us to incorporate
regions with different reactivity, for instance to account for changes
in temperature or different types of substrates.

All time derivatives are discretized using an implicit Euler method,
whereas linear interpolation is used to approximate values
at cell faces. The use of implicit methods ensures that fractional surface coverages remain bounded between 0 and 1. The
solution of the resulting system of equations is solved
using OpenFOAM's built-in algorithms. In particular,
we used OpenFOAM's preconditioner
biconjugate gradient (PBiCG) solver, a standard Krylov subspace solver that allows the use of a runtime selectable preconditioner. For this work, we used the simplified diagonal-based incomplete LU (DILU) preconditioner.

The velocity field $\bm{u}$ used as input for the advection, diffusion, reaction
equation of gas-phase species (Eq. \ref{eq_ci}) has been calculated using OpenFOAM's implementation
of the SIMPLE algorithm for the Navier-Stokes equations.\cite{Ferziger2002}
The velocity fields have been obtained assuming laminar flow conditions,
which is a reasonable assumption for the low Reynolds
numbers involved in the low pressure reactors considered in this work.
We have also used non-slip boundary conditions for the flow velocity,
which are consistent with the low Knudsen numbers in our experimental
condition ($\mathrm{Kn} < 0.01$, assuming a mean free path of 50 microns at 1 Torr and characteristic reactor width of the order of 1 cm).
OpenFOAM's
current implementation naturally allows us to extend the simulation conditions to
non-isothermal conditions. We have used these in the past to model other
configurations, such as vertical MOCVD reactors, with strong thermal
gradients.\cite{Ju_MOCVD_2017} Likewise,
it is possible to generalize the model to consider turbulent flow conditions. Both
conditions are outside the scope of the present work. In particular, for turbulent
flows one needs to consider the effect of turbulent transport, which becomes
the dominant mechanism of precursor mixing.

The code has been run both in
off-the-shelf desktops and laptops and in Blues, one of the clusters at
Argonne's Laboratory Computing Research Center. This
allowed us to explore process parameters in a massively parallel fashion.

\subsection{Transport coefficients}

The model described above depends on the
values of the kinematic viscosity $\nu = \mu/\rho$
of the carrier gas as well as the pair diffusivities
of the different species in the carrier gas $D_i$.

We have used the Chapman-Engskop expression
for the pair diffusivity:\cite{Rosner}
\begin{equation}
    D_{ij} = \frac{3k_BT}{8p} \sqrt{\frac{k_BT}{2\pi} \frac{M_i+M_j}{M_iM_j}}\frac{1}{\sigma_{ij}^2 \Omega_D(k_BT/\varepsilon_{ij})}
\end{equation}

Where $\sigma_{ij}$ and $\varepsilon_{ij}$ are coefficients for the pair potential, which is assumed to be spherically symmetric. In the case
of a Lennard-Jones model, $\Omega_D(k_BT/\varepsilon_{ij})$ is a function
that can be parametrized as:
\begin{equation}
    \Omega_D(k_BT/\varepsilon_{ij}) \approx 1.22(k_BT/\varepsilon_{ij})^{-0.16}
\end{equation}

\section{Results}

\begin{figure}
\includegraphics[width=3.5in]{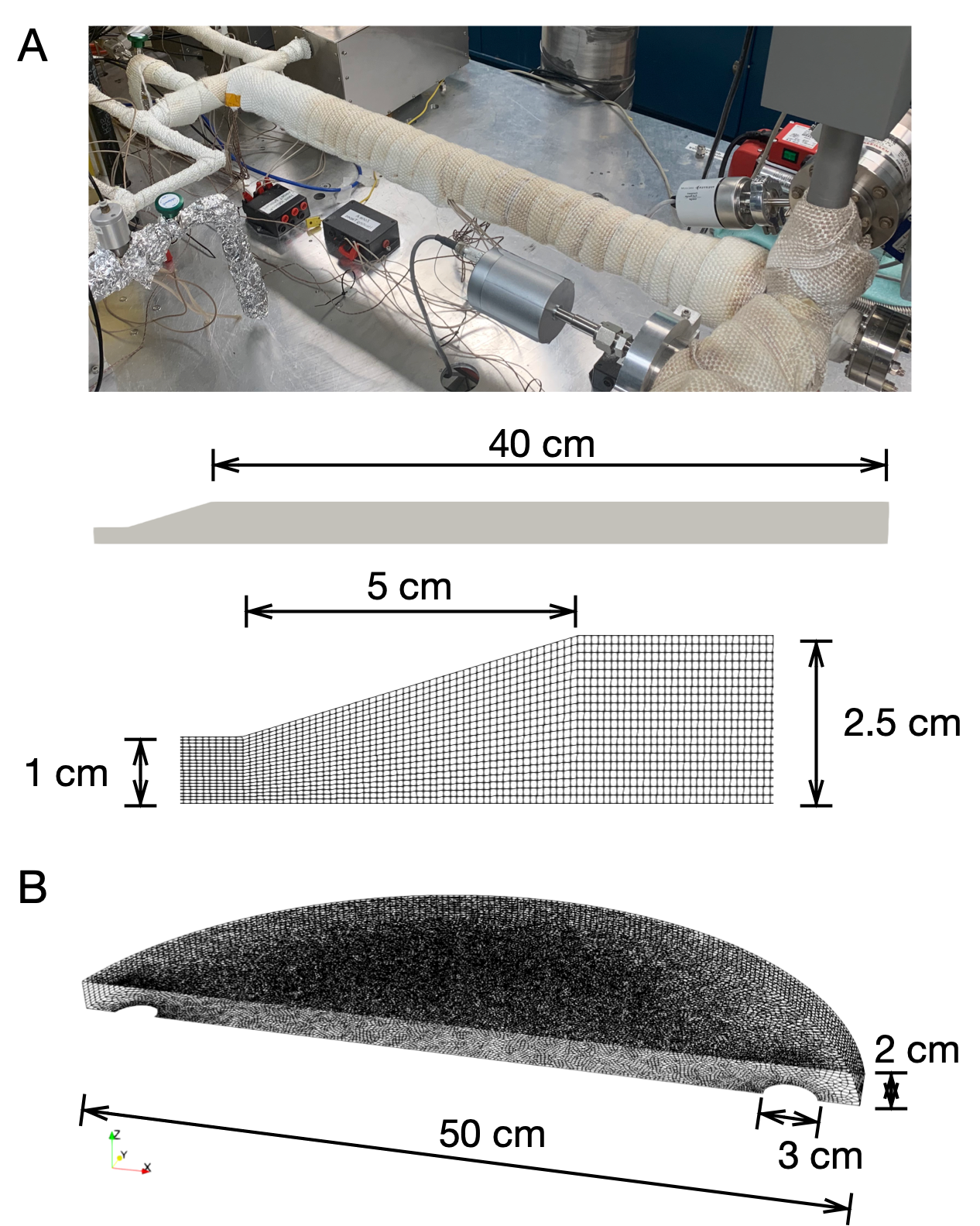}
\caption{\label{fig_mesh} (A) Simulation domain for a horizontal cross-flow reactor. The 2D domain has cylindrical symmetry and incorporates an expanding section bridging
the inlet manifold with the reactor tube. A picture of our experimental setup is
shown for comparison. (B) Simulation domain for a 300mm wafer reactor. The simulation
domain comprises a circular disk with a 50 cm diameter and 3 cm diameter inlet and
outlet regions. The height of the reactor is 2 cm. A 300 mm wafer region is
placed at the center of the reactor and treated as a separate area.
Only half of the reactor is modeled, with
mirror boundary conditions used at the bisecting plane.}
\end{figure}

\subsection{Simulation domains and carrier gas flow}

In this work, we have considered two main reactor geometries shown in Figure 
\ref{fig_mesh}. The first
geometry is a 2D model of a 5 cm diameter cross-flow cylindrical tube reactor.\cite{Elam_viscousflow_2002}
This model reproduces the geometry
of three of the experimental reactors in our laboratory, one of which is
shown in Figure \ref{fig_mesh}(A). We have created an axisymmetric mesh using OpenFOAM's mesh generation utility blockMesh. The mesh used in this work is composed of 9,400 cells, with a spatial resolution in the downstream
direction of 1 mm. Meshes with 2 times higher spatial resolution
were also created to confirm that the simulation results are independent
of mesh size.

The second geometry corresponds to a large area cross-flow horizontal
circular reactor for 300 mm wafers with 30 mm circular inlet and outlet and a total diameter of 50 cm. The reactor is 2 cm tall [Fig. \ref{fig_mesh}(B)]. The mesh
has been generated using Gmsh,\cite{Geuzaine_Gmsh_2009} an Open Source mesh generator, and then converted to an OpenFOAM-compatible
format using the utility gmshToFOAM. The mesh used in this work is
composed of 148,390 cells. Again, meshes with 2 times higher spatial resolution
were also created to confirm that the results were independent of
the mesh size. 

\begin{table}
\caption{\label{tab_conditions}
Summary of parameters used in the simulations
}
\begin{ruledtabular}
\begin{tabular}{lccc}
\textrm{Parameter}&
\textrm{Symbol}&
\textrm{Value}&
\textrm{Units} \\
\colrule
\textrm{Precursor diffusivity} & $D_1$ & 0.01 & m$^2$s$^{-1}$\\
\textrm{Kinematic viscosity} & $\nu$ & 0.05 & m/s \\
\textrm{Process temperature} &  $T$ & 473 & K \\
\textrm{Precursor molecular mass} & $M_1$ & 150 & amu \\
\textrm{Surface site area} & $s_0$ & 24 & nm$^{2}$ \\
\textrm{Byproduct diffusivity} & $D_{bp}$ & 0.05 & m$^2$s$^{-1}$\\
\end{tabular}
\end{ruledtabular}
\end{table}

\begin{figure}
\includegraphics[width=3in]{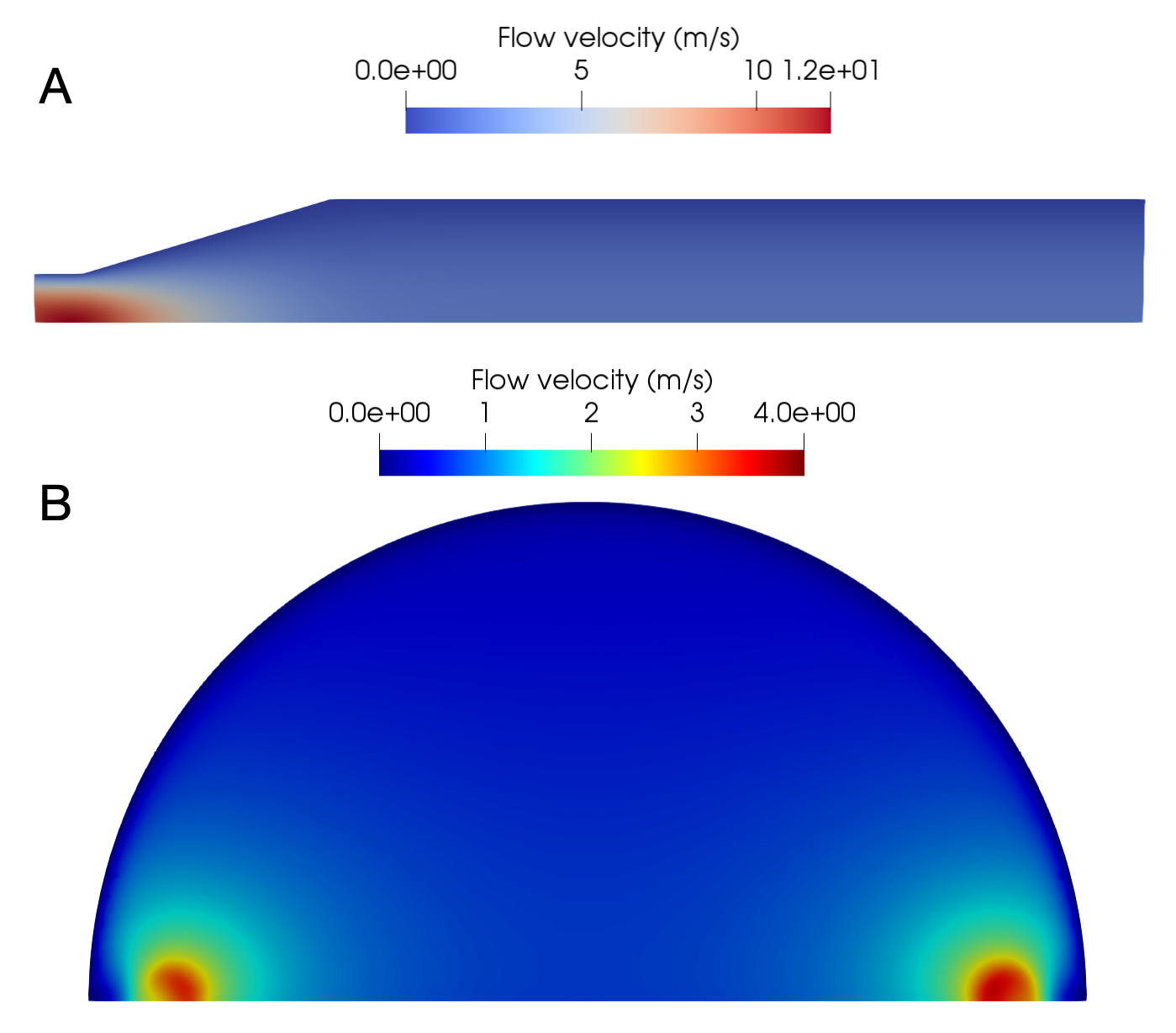}
\caption{\label{fig_velocity} Velocity profiles for our two simulation
domains: (A) tube cross flow geometry (B) 300 mm wafer reactor.}
\end{figure}

The same flow conditions were used for all simulations in this work.
For the tube geometry, we adapted the inlet velocity to ensure
that the average flow velocity inside the reactor was 1 m/s, which
is in agreement with the values expected in our experimental setup,  and considered a
 pressure of 1 Torr.
In Figure \ref{fig_velocity}(A), we show the steady state magnitude
value of the flow velocity as it transitions from the inlet to
the wider reactor region. The kinematic viscosity (Table \ref{tab_conditions}) is high enough
to ensure that viscous terms compensate the inertial term and the
flow quickly becomes fully developed, opening up in the reactor area
without 
any wall separation that may lead to recirculation patterns and stagnation
points at the inlet.

In the case of the 300 mm wafer reactor, the velocity field was
obtained assuming 300 sccm of total flow at a base pressure of 1 Torr.
In Figure \ref{fig_velocity}(B) we show the magnitude of the flow velocity
at a cross section located at mid height of the reactor.
Both are consistent with stable flow patterns, with an
average velocity of 0.5 m/s over the wafer area.

\subsection{Ideal ALD process}

In this Section we first focus on the ideal ALD model given in
Section \ref{sec_ideal}. Unless explicitly mentioned, the parameters used
for the simulations are given in Table \ref{tab_conditions}.

\subsubsection{Benchmark with plug flow model}

In a previous work, we considered a plug flow approximation
of a cross-flow reactor and derived analytic solutions for
the coverage profiles for the ideal ALD process.\cite{YanguasGil2014}
In terms of the precursor pressure $p_0$, the average 
flow velocity $u$, dose time $t_d$ and the same parameters
used in Eq. \ref{eq_ideal}, the predicted saturation coverage as
a function of the dose time was given by the following expression:
\begin{equation}
    \label{eq_ald}
    \Theta(z;t_d) = \frac{e^{t_d/\bar{t}} -1}{e^{z/\bar{z}} + e^{t_d/\bar{t}} -1}
\end{equation}
where:
\begin{equation}
    \bar{t} = \frac{4k_BT}{s_0 v_{th} \beta_0 p_0} 
\end{equation}
and
\begin{equation}
    \label{eq_z}
    \bar{z} = \frac{V}{S}\frac{4 u}{v_{th} \beta_0}
\end{equation}
Here, the volume to surface ratio $\frac{V}{S}$ is equal
to $d/2$ for a parallel plate reactor and $R/2$ for a 
tubular reactor with $R$ the radius of the cylindrical
tube.

In that same work we also showed that Eq. \ref{eq_ald}
yielded excellent
agreement with experimental saturation profiles obtained during
the ALD of aluminum oxide from trimethylaluminum (TMA) and water.\cite{YanguasGil2014}

\begin{figure}
\includegraphics[width=3in]{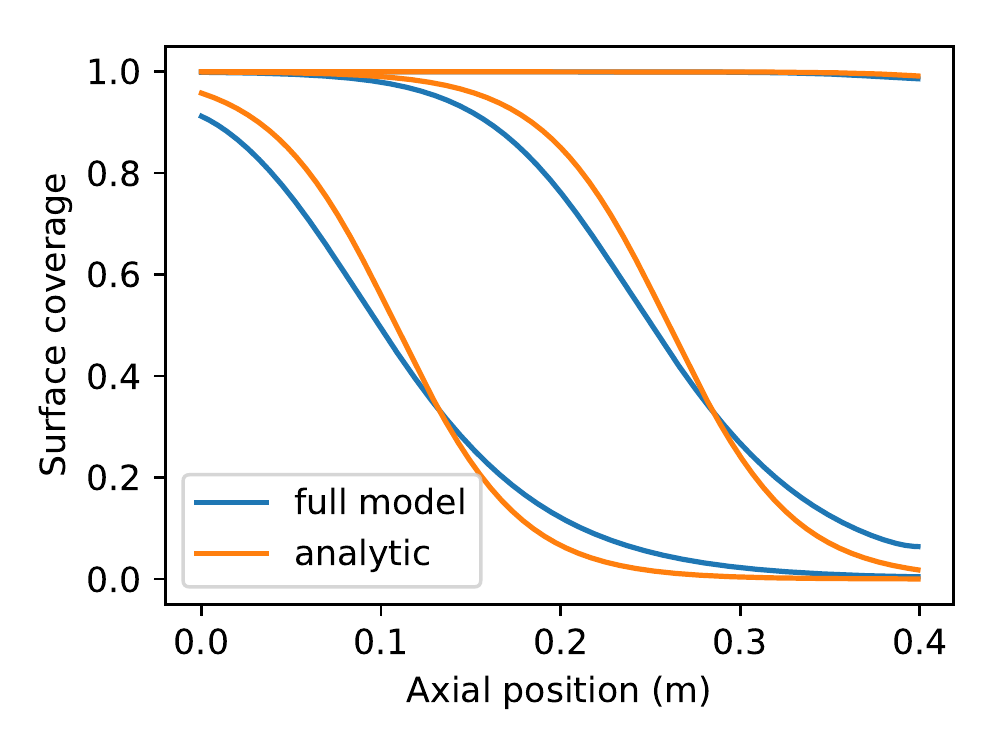}
\caption{\label{fig_bench} Comparison between the full 2D simulation
and the analytic expression under the plug-flow approximation (Eq. \ref{eq_ald}) derived in Ref. \cite{YanguasGil2014}.
The results were obtained for three
undersaturated precursor doses of 0.05, 0.1, and 0.2 s, an average precursor pressure at the inlet of 20 mTorr, and a sticking
probability $\beta_{10} = 10^{-2}$}
\end{figure}

We can use our 2D simulations of our tubular cross-flow reactor
to establish a comparison between the model developed in this
work and Eq. \ref{eq_ald}. Such a comparison is shown in
Figure \ref{fig_bench}, where we present simulated growth profiles
along our reactor for three different dose times: 0.05 s, 
0.1 s and 0.2 s, where in all cases the average precursor
pressure at the inlet $p_0$ is set to 20 mTorr. The results,
which have been obtained for a reaction probability $\beta_{10}=10^{-2}$, show a good agreement between this
work and the analytic expression.
A key difference between this work and the plug flow approximation
is that the latter neglects the presence of axial and radial diffusion: the
combined effect of these two factors is a softening of the
growth profiles with respect to the plug flow model.

Since our simulation domain incorporates part of the reactor
inlet, in order to produce Fig \ref{fig_bench} we had to 
take into account the effect of upstream consumption in Eq. \ref{eq_ald}.
We did so by considering an offset in the axial coordinate given by:
\begin{equation}
    \label{eq_upstream}
    \Delta z = \int_0^l z \frac{R(z)}{R_0} dz
\end{equation}
where $l$ is the length of the inlet, $R(z)$ is the radius at that
specific location, and $R_0$ is the radius of the reactor tube. This expression
is the result of considering the spatial dependence of $\bar{z}$ (Eq. \ref{eq_z}) in the inlet as the radius (and therefore the mean
flow velocity) changes with respect to the radius of the reactor
tube.

\subsubsection{Impact of reaction probability}

\begin{figure}
\includegraphics[width=3in]{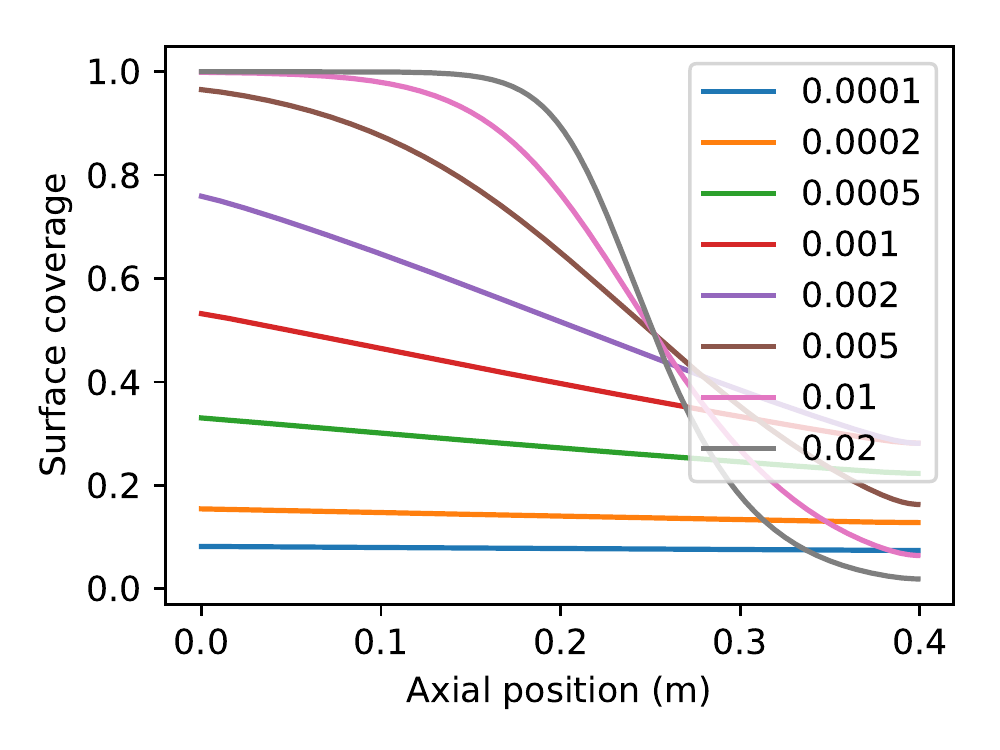}
\caption{\label{fig_beta} Growth profiles for a 200 ms and 20 mTorr precursor dose for different values of the reaction probability
$\beta_{10}$. Higher reaction probabilities lead to steeper growth
profiles.}
\end{figure}

The reaction probability $\beta_{10}$ is by far the most influential
factor in the ideal ALD model. In Figure \ref{fig_beta}, we show
the impact of this parameter in the reactor growth profiles:
all curves in Fig. \ref{fig_beta} have been obtained
considering the same dose time (0.2 s) and precursor pressure (20 mTorr) and different values of $\beta_{10}$. As shown in 
previous works, a decrease of the reaction probability decreases
the steepness of unsaturated growth profiles in the reactor. It
is important to note, though, that a reduction of one order
of magnitude in the value of $\beta_{10}$ does not necessarily
reduce the total mass uptake by an order of magnitude. Instead,
if the reaction probability is high enough, a reduction in
the reaction probability simply redistributes the way
the mass is deposited in the reactor. Only when the reaction
probability is low enough, the system transitions from a 
transport-limited to a reaction-limited regime, and the mass
uptake is directly proportional to $\beta_{10}$

\begin{figure}
\includegraphics[width=3in]{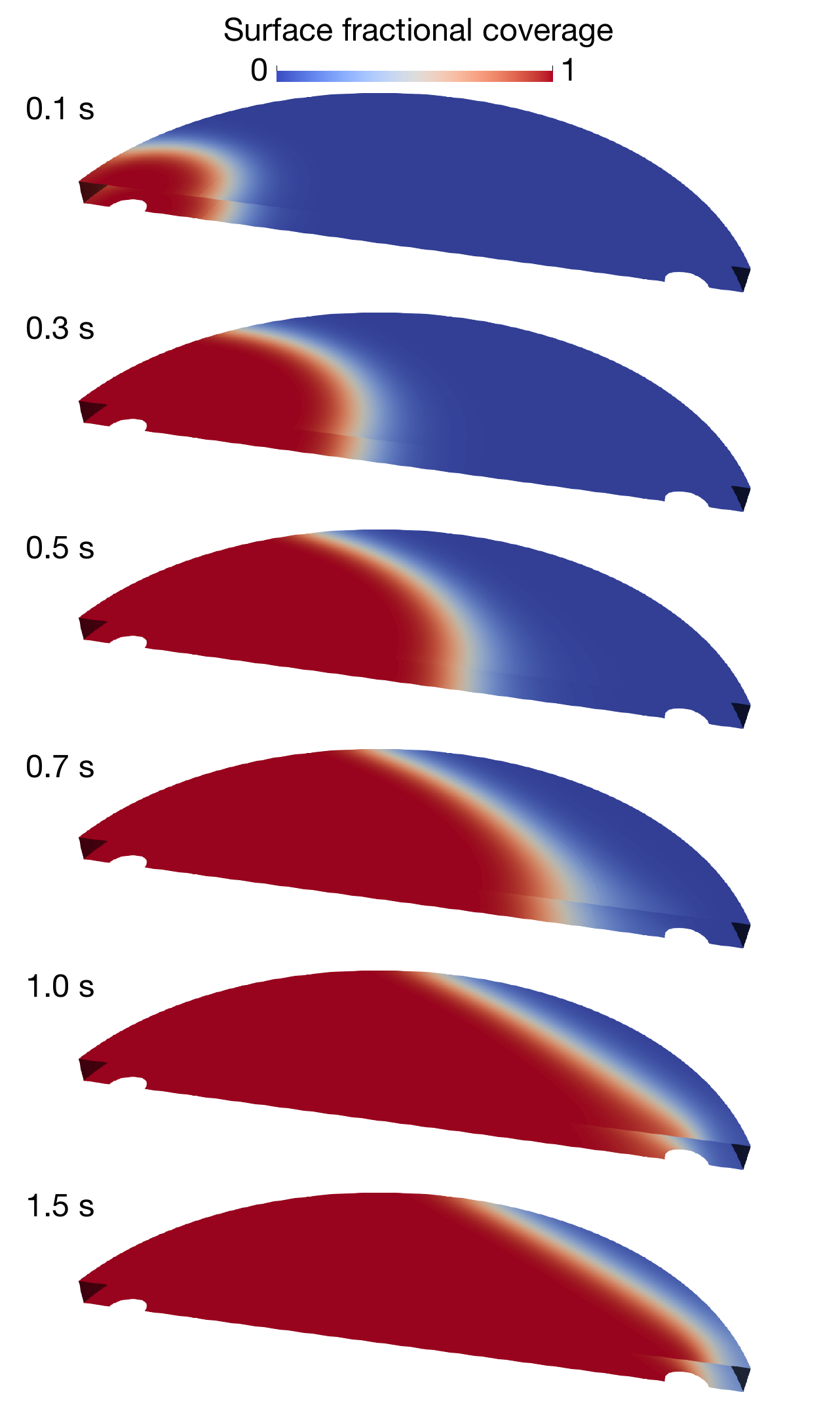}
\caption{\label{fig_time}Evolution
of surface coverage as a function of time during a 0.3 s dose and 1 s purge. The simulations assume a precursor pressure at the inlet of $p_0=75$ mTorr, and a reaction probability for the self-limited process
$\beta_{10}=10^{-2}$}
\end{figure}

The same trends are reproduced when instead of the tubular reactor we consider the 300 mm wafer reactor. In Figure \ref{fig_time}
we show the evolution of surface coverage during a single
half cycle comprising a 0.3 s dose and 1 s purge. The simulations
assume a precursor pressure of $75$ mTorr and $\beta_{10}=10^{-2}$.
The evolution of the surface coverage follows the steep
saturation profile that is expected from high sticking
probability processes.

\begin{figure}
\includegraphics[width=3in]{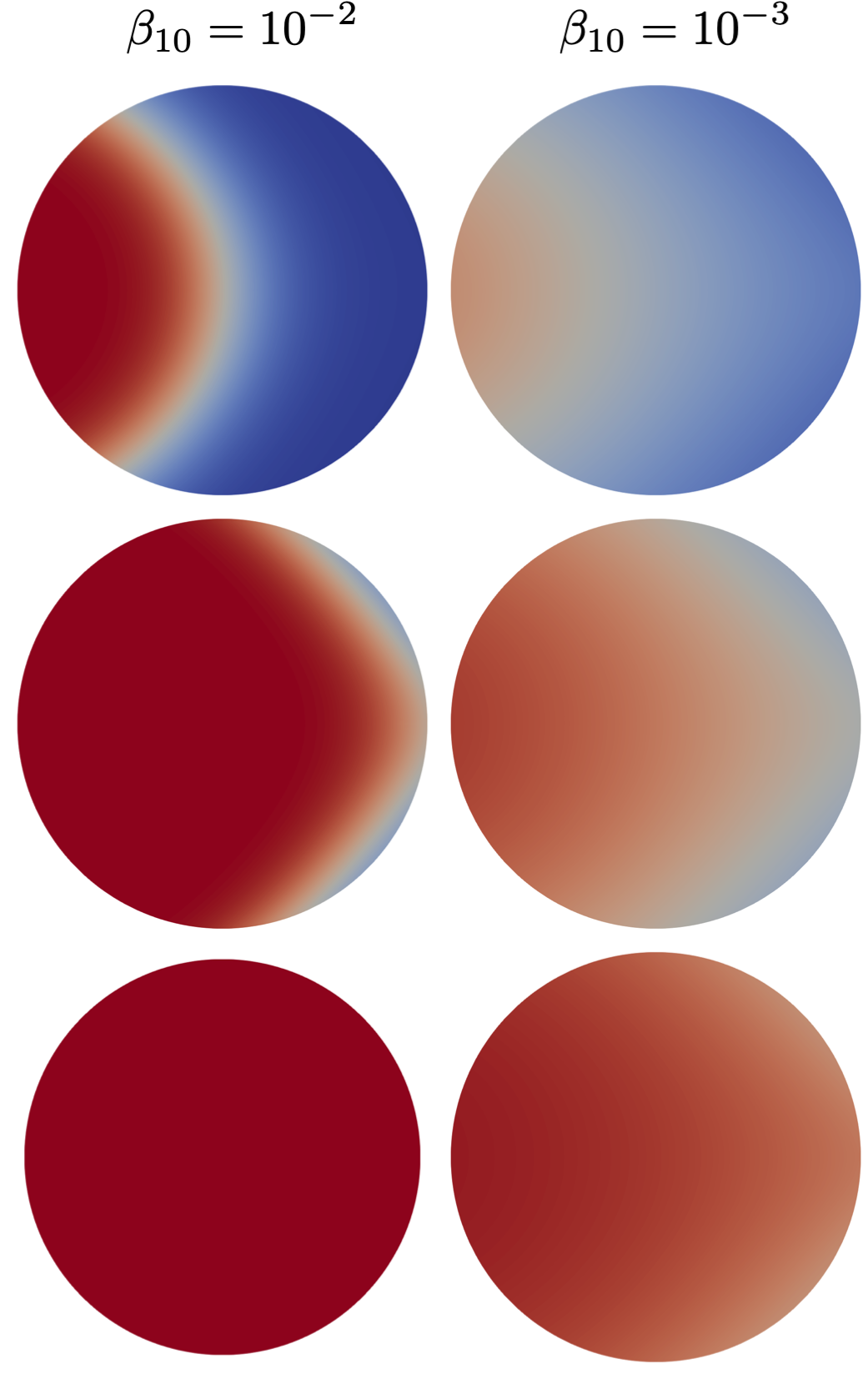}
\caption{\label{fig_wafer0}Surface coverage for increasing dose times of 0.1, 0.2, and 0.4 s. The simulations assume a precursor pressure at the inlet of $p_0=75$ mTorr, and a reaction probability for the self-limited process are $\beta_{10}=10^{-2}$ (left) and
$\beta_{10}=10^{-3}$ (right)}
\end{figure}

In Figure \ref{fig_wafer0}, we show the coverage profiles on 30 cm wafer substrates for increasing dose times and two values of the reaction probability: $\beta_{10}=10^{-2}$ and $\beta_{10}=10^{-3}$.
The results show the smoothening of the profiles and the
increase in saturation times as the reaction probability
decreases.

\subsubsection{Modeling in-situ QCM and QMS}

As shown in the previous section, undersaturated growth profiles
carry out information about the underlying kinetics of
self-limited processes. However, the information that they
provide is static, representing the cumulative effect of
a whole dose. In contrast, in-situ techniques such
as quartz crystal microbalance (QCM) and quadrupole
mass spectrometry (QMS) have enough temporal resolution to
provide mechanistic information during each dose. One
challenge of extracting kinetic data from these
techniques is that kinetic effects are convoluted with
the reactive transport of the precursor inside the
reactor. Here we focus on simulating the output of
these two techniques in order to understand how the
flow and process conditions affect the measurements.

\begin{figure}
\includegraphics[width=3in]{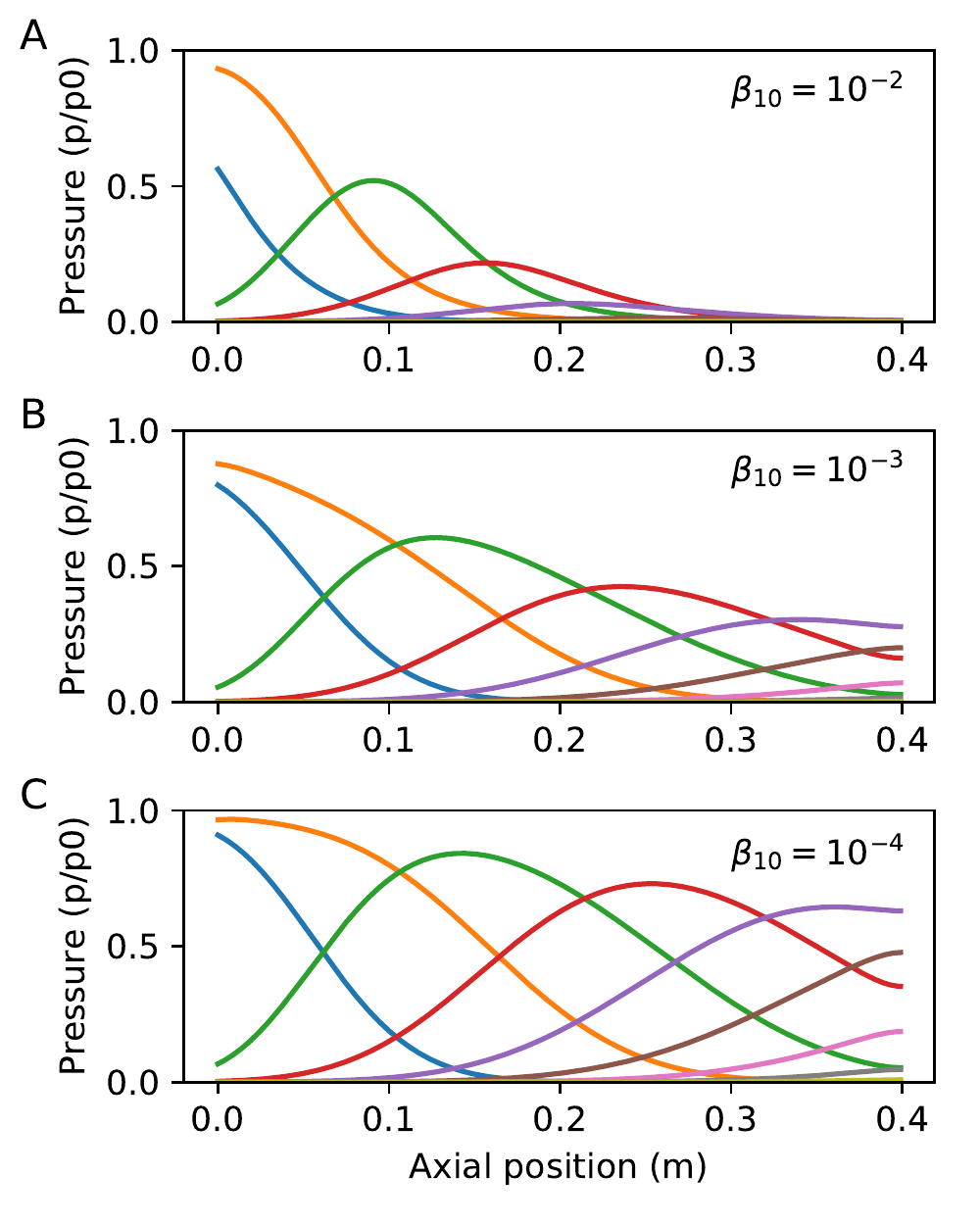}
\caption{\label{fig_density} Precursor pressure inside the
reactor during a single under-saturated dose (0.2 s at 20 mTorr). 
Snapshots are taken at 0.1s intervals: (A) $\beta_{10}=10^{-2}$; 
(B) $\beta_{10}=10^{-3}$; (C) $\beta_{10}=10^{-4}$}
\end{figure}

The transport of the precursor and the
reaction byproducts is strongly influenced by the
reaction probability $\beta_{10}$. In Figure
\ref{fig_density} we show the the precursor partial
pressure inside our tube reactor model at different
snapshots in time taken at 0.1 s intervals during
a single saturated dose. In particular, we
compare processes with three different values of
the reaction probability: $10^{-2}$, $10^{-3}$,
and $10^{-4}$. In the high reaction probability case,
the precursor is fully consumed before it reaches
the end of the reactor. However, as the reaction probability
decreases, an increasing fraction of the precursor reaches
the outlet.

\begin{figure}
\includegraphics[width=3in]{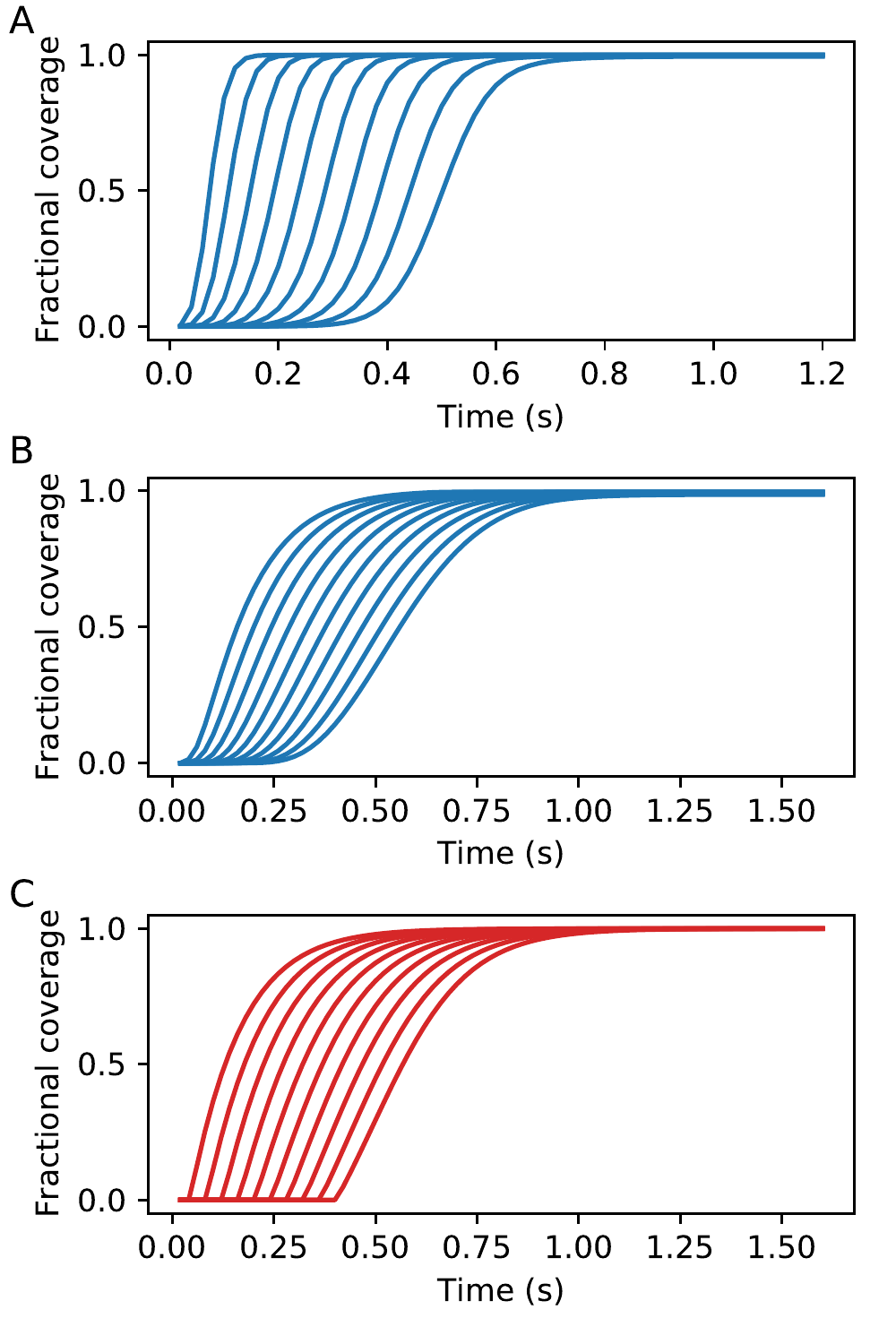}
\caption{\label{fig_qcm} Simulated QCM traces showing the evolution of the fractional surface coverage during a single saturating dose at 10 different reaction positions located at a 4 cm distance from each other: (A) $\beta_1 = 10^{-2}$ (B) $\beta_2=10^{-3}$. In both cases the precursor pressure is 50 mTorr. (C) Plug flow approximation
(Eq. \ref{eq_qcm}) using the same conditions as in (B) }
\end{figure}

This behavior impacts the shape of the QCM profiles depending
on their position inside the reactor. In Figure \ref{fig_qcm}
we have considered the evolution of the fractional surface
coverage during a single saturating dose at 10 different
positions in our tubular reactor, each placed 4 cm apart. Under
the ideal ALD conditions considered in this section, there is
a one-to-one correspondence between the fractional surface
coverage and the mass gain observed by the QCM.

The results in Figure \ref{fig_qcm} show that
time delays of up to 0.5 seconds
can be expected between the first and the last QCM of 
the array. Precursor transport is a key factor in this offset,
with an small contribution due to the influence of
upstream precursor consumption. Upstream precursor consumption
also leads to a change in slope observed as the QCM moves
further from the inlet. This hinders our ability
to extract quantitative information from data coming
from a single QCM within
a single half-cycle of a self-limited process without 
additional flow information. 

When we compared the growth profiles obtained with the
full simulation with the 1D plug flow model we observed a
good agreement between the final saturation profiles predicted
by both models. It is therefore reasonable to try to account
for flow effects in Figure \ref{fig_qcm} using  Eq. \ref{eq_ald} 
together with Eq. \ref{eq_upstream}. The plug flow
model predicts a mass gain as a function
of time for a QCM located at a position $z$ given by:
\begin{equation}
    \label{eq_qcm}
    \Theta(z, t) = \frac{e^{[t-\Delta t]^+/\bar{t}}-1}
        {e^{[t-\Delta t]^+/\bar{t}}+e^{z/\bar{z}}-1}
\end{equation}
where $[\cdot]^+$ is a rectifying function that is zero whenever
the argument is negative. The value of $\Delta t$ represents the
delay in the arrival of the pulse, which in the plug flow approximation
is simply given by:
\begin{equation}
    \Delta t = \frac{z}{u}
\end{equation}
where $u$ is the average flow velocity in the tube. In Figure \ref{fig_qcm}(C) we show the predicted values of fractional coverage with time under  the same conditions of \ref{fig_qcm}(B). While the agreement is not perfect, Eq. \ref{eq_qcm} provides a good approximation that could be used  to discriminate between ideal and non-ideal self-limited processes from QCM data.

The reaction
probability does have a strong impact on the distribution of
mass gains that would be observed using more than
one sensor at a given time: for a
high sticking probability process, mass gains sharply transition from fully
saturated values in the upstream sensors to zero mass gain in the
doenstream sensors [Fig. \ref{fig_qcm}(A)], whereas for lower
values of the reaction probability [Fig. \ref{fig_qcm}(B)] the
expected difference in mass gain between upstream and donwstream sensors
is expected to be smaller.

\begin{figure}
\includegraphics[width=3in]{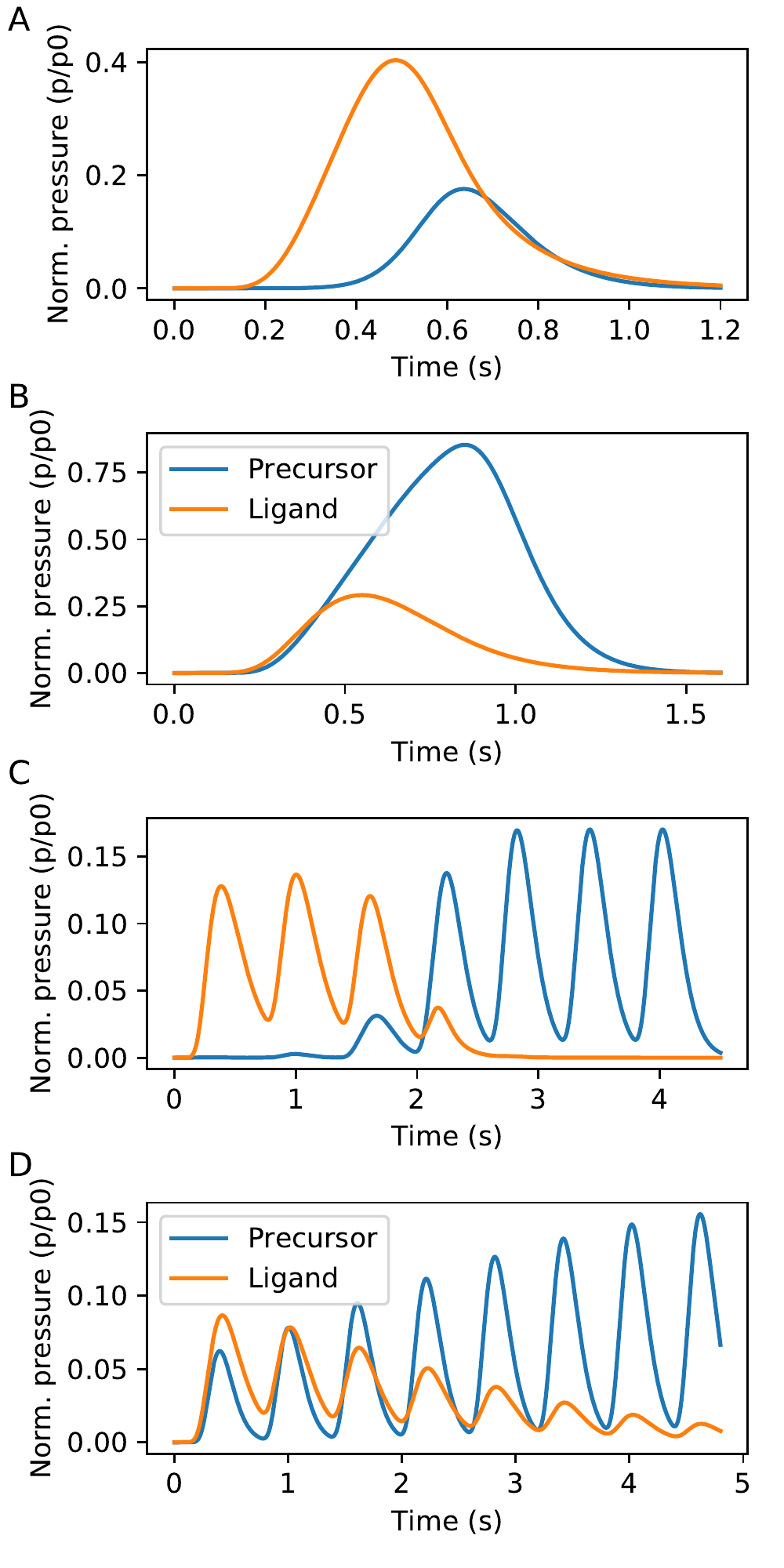}
\caption{\label{fig_downstream} Normalized pressure of precursors
and ligand species at the downstream position of
the tubular reactor: (A) Single dose, $t_d=0.2$ s, $\beta_1 = 10^{-2}$ (B) Single dose, $t_d=0.4$ s, $\beta_1=10^{-3}$. (C) Microdose sequence, $\beta_{01} = 10^{-2}$ (D) Microdose sequence $\beta_{02}=10^{-3}$. A high reaction probability leads to a temporal separation between the precursor and ligand peaks. In all cases the
average precursor pressure at the inlet is $p_0=50$ mTorr.}
\end{figure}

In order to model QMS input, we have tracked as a function of
time the partial pressures of both the precursor and byproduct
species at a single location at the downstream position
of the reactor. This is a common experimental configuration.
In Figure \ref{fig_downstream} we show the concentration of
both species for the same two processes used for the QCM
analysis: the reaction probability has a strong impact on the temporal
separation of the traces coming from the precursor and the byproduct.
This is a natural consequence of the results shown in Figure \ref{fig_density}, where for high enough reaction probabilities all the precursor initially reacts inside the reactor. At the
same time, reaction byproducts are released since the beginning
of the dose. This creates an offset between the two contributions. As the reaction probability goes down, this offset is reduced,
as clearly seen in Figure \ref{fig_downstream}(B) for the case
of $\beta_{10}=10^{-3}$.

This separation becomes more apparent if we break down a single dose into a sequence of microdoses. In Figure \ref{fig_downstream}(C) and \ref{fig_downstream}(D) we show the precursor and byproduct partial pressures during a sequence of microdoses, where each pulse corresponds to a 0.1 s dose. In the high sticking probability case [Fig. \ref{fig_downstream}(C)] the first pulses are dominated by the contribution coming from the reaction byproducts,
and the transition between pure byproduct and pure precursor signals
occurs rather abruptly at ~2 s. In contrast, a lower sticking probability
[Fig. \ref{fig_downstream}(D)] leads to byproduct and precursor signals
that persist throughout the microdose sequence and a gradual transition
between byproduct and precursor signals.

\subsection{Non-ideal self-limited processes}

\subsubsection{Soft-saturating processes}

The first generalization of the ideal self-limited model
that we have considered is the presence of more than one reactive pathway
on the surface: we have incorporated 
a second self-limited reaction pathway, comprising
a fraction $f$ of the surface sites and characterized by
a different reaction probability $\beta_{1b}$ (Section \ref{sec_softsat}).
This allows us to model
soft-saturating processes with fast initial rise and slow saturation.

\begin{figure}
\includegraphics[width=3.5in]{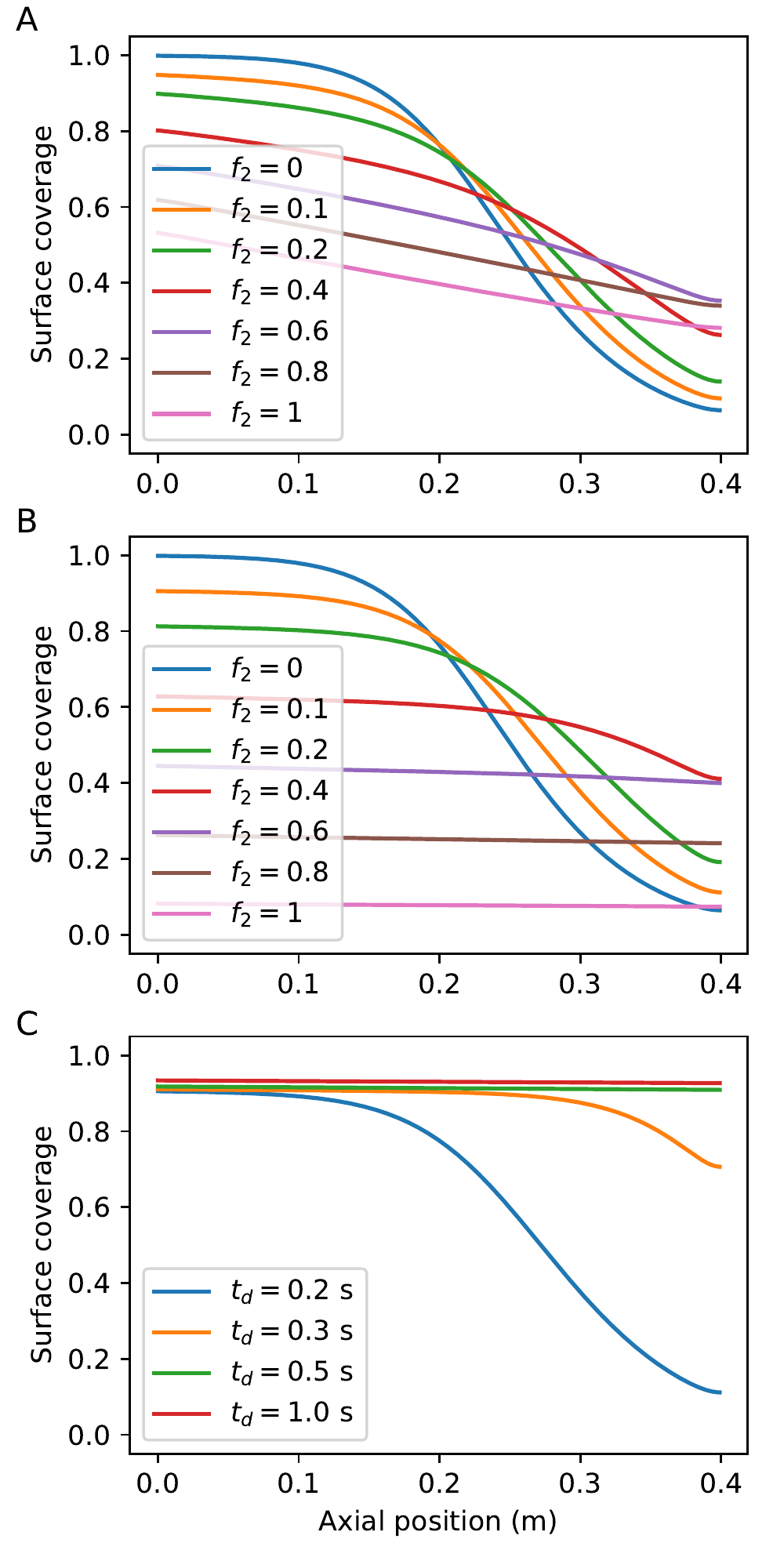}
\caption{\label{fig_softprofs} Reactor growth profiles for a soft-saturating self-limited model with two components (A) $\beta_{1a} = 10^{-2}$ and $\beta_{1b}=10^{-3}$ (B) $\beta_{1a} = 10^{-2}$ and $\beta_{1b}=10^{-4}$. All profiles correspond to the same unsaturated dose time of 0.2 s for an average precursor pressure at the
inlet of 50 mTorr.}
\end{figure}

In Figure \ref{fig_softprofs} we show growth profiles in our tube reactor
configuration for
a self-limited process with a fast and a slow component for varying fractions of
the slow component, $f$. The growth profiles have been obtained using
the same dose times (0.2 s) and precursor pressure in the inlet (50 mTorr)
for all the cases. In
Figure \ref{fig_softprofs}(A), the fast and slow components
have reaction probabilities of $10^{-2}$ and $10^{-3}$, and they
show the progressive transition from a more step-like saturation profiles
at $f=0$ to a more gentle, almost linear profile for $f=1$. 

In Figure \ref{fig_softprofs}(B) the probability of the slow component is two orders of magnitude smaller, $\beta_{1b} = 10^{-4}$. The interesting aspect in this case is that
for small values of $f$, the region closer to the inlet shows the type of plateau
that would be expected from a fully saturated process, except at surface
coverages that are well below saturation. In Figure \ref{fig_softprofs}(C) we
show growth profiles for increasing dose times for the case of $f=0.2$ and
$\beta_{1b} = 10^{-4}$. After a dose time of 0.5 s, the process has the appearance
of saturation but its slow component is still evolving, resulting  
on a slightly higher coverage at 1s dose time. The slow self-limited component could be easily mistaken for
a non self-limited component if the timescales for saturation are large enough.
If we compare this result
with the literature, there are several examples of this behavior in the case of ALD:
for instance, it has been shown that extremely large doses of water can lead
to slightly larger growth per cycles than the conventional ALD process.\cite{Matero_water_2000}

\begin{figure}
\includegraphics[width=3in]{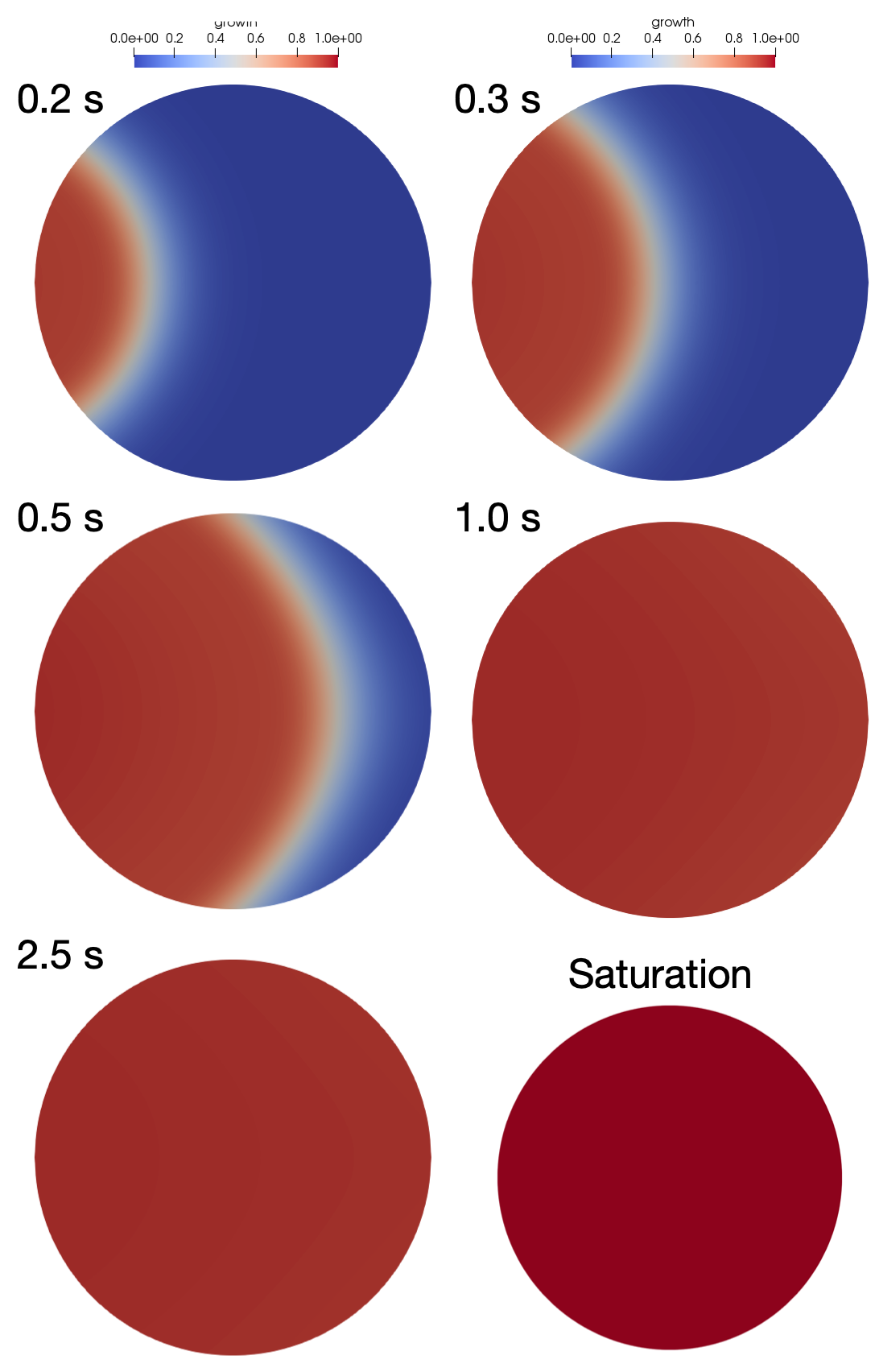}
\caption{\label{fig_softwafers} Evolution of coverage as a function of
time for a self-saturating process where $f=0.1$ and $\beta_{1b}=10^{-4}$. While
the fast component reaches saturation after 0.5 s, the remaining 10\% of the
sites require an order of magnitude higher dose times, resulting in a homogeneous
yet unsaturated profile. A
fully saturated wafer from an ideal self-limited process is shown as comparison.}
\end{figure}

If we now consider this model on the 300 mm wafer reactor, we observe similar trends: in Figure \ref{fig_softwafers} we show the surface coverage on 300 mm wafers for increasing
values of dose times for the same soft saturating process represented
in Figure \ref{fig_softprofs}(C). After 1 s, the whole wafer has almost identical surface coverage, yet it is still 10\% below its saturation value.

\begin{figure}
\includegraphics[width=3.5in]{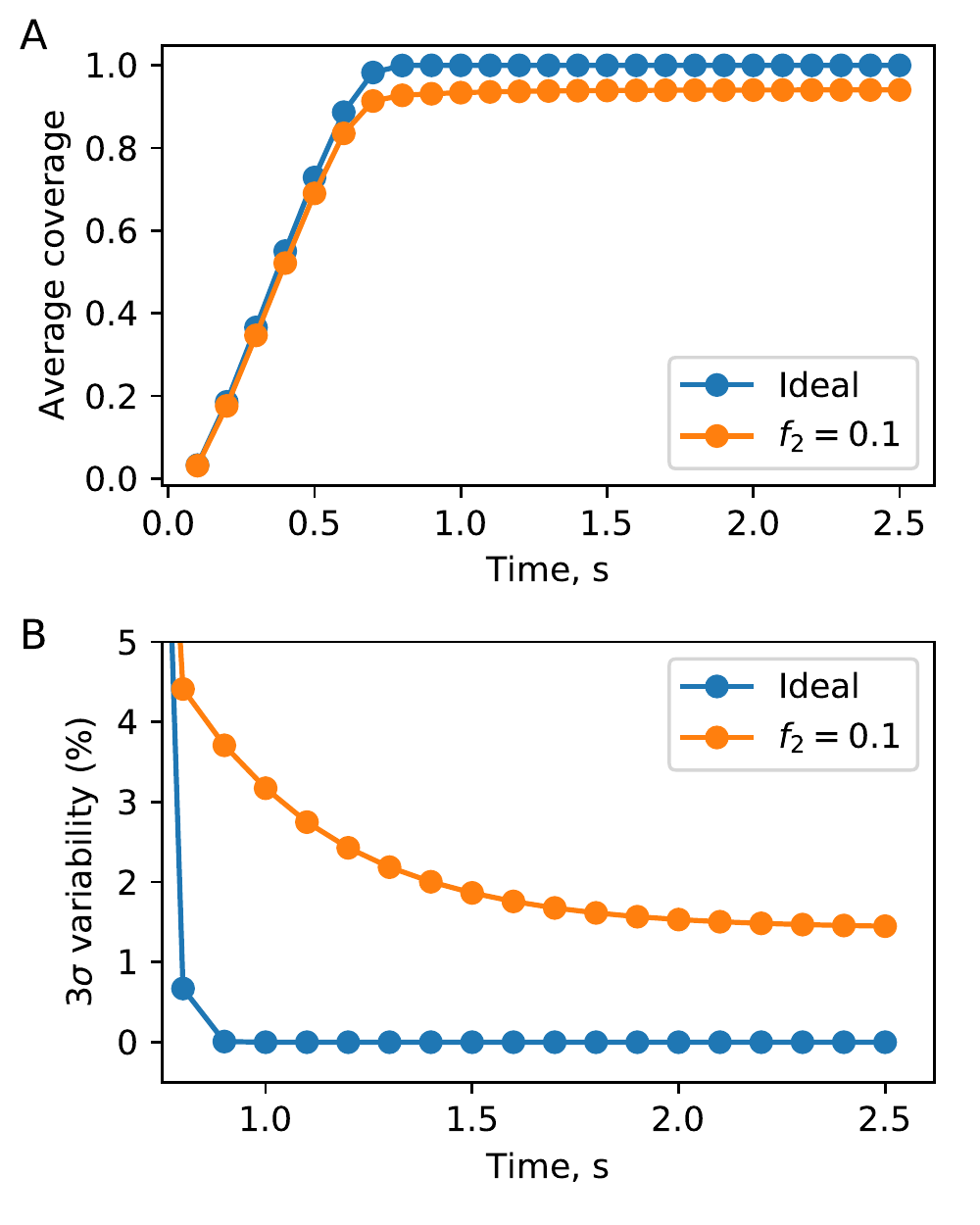}
\caption{\label{fig_softrms} (A) Average wafer coverage and (B) $3\sigma$ wafer thickness variability for an ideal and a soft-saturating ALD process. Simulation conditions are
the same as for Figure \ref{fig_softprofs}(C)}
\end{figure}

A second consequence of this soft-saturating behavior is a process
that appears to have reached saturation but that still
has a measurable variability inside the reactor. Using
the $3\sigma$ standard deviation of surface coverage 
as a measure of within-wafer variability, in Figure \ref{fig_softrms}
we compare the saturation curve of 
the soft-saturating process with that of the ideal process: 
both curves are very similar, except that the soft-saturating case
"saturates" at a lower value of the fractional coverage [Fig. \ref{fig_softrms}(A)]. On the other hand, the $3\sigma$ variability
of the coverage across the wafer is much higher in the
soft-saturating case and decreases asymptotically
with dose times. 

\subsubsection{Site blocking by ligands and reaction byproducts}

A second generalization of the ideal self-limited interactions
considers the effect of ligands or reaction byproducts competing
with the precursor for adsorption sites. This effect has
been well documented in the ALD literature, leading to self-limited
yet inhomogeneus growth profiles, as it is for instance the case
of the ALD of TiO$_2$ from titanium tetraisopropoxide and water and
titanium tetrachloride and water.\cite{Ritala_TTIP_1993, Elers_uniformityreview_2006} Growth modulation
studies in ALD showed that alkoxides, betadiketones, and carboxylic acids
are some of the moieties that interfere with the adsorption of
ALD precursors.\cite{YanguasGil2013}

As described in Section \ref{sec_site}, we have modeled 
the presence of competitive adsorption between precursor molecules
and reaction byproducts by considering that byproducts
can irreversibly react with surface sites through
a first order Langmuir kinetics characterized by a reaction probability
$\beta_{bp0}$. Under this assumption,
the surface now is composed of two types of
adsorbed species, and we can therefore define a
precursor fractional coverage $\Theta$ and a byproduct
fractional coverage $\Theta_{bp}$.

The sticking probability of both the precursor and reaction
byproducts is  proportional to
the fraction of available sites:
\begin{eqnarray}
    \beta_1 & = & \beta_{10}\left( 1 - \Theta_1 - \Theta_{bp}\right)\\
    \beta_{bp} & = & \beta_{bp0}\left( 1 - \Theta_1 - \Theta_{bp}\right)\\
\end{eqnarray}
Consequently, both species are competing for the same pool of
surface sites. The surface becomes
unreactive when $\Theta + \Theta_{bp} = 1$.

\begin{figure}
\includegraphics[width=3.5in]{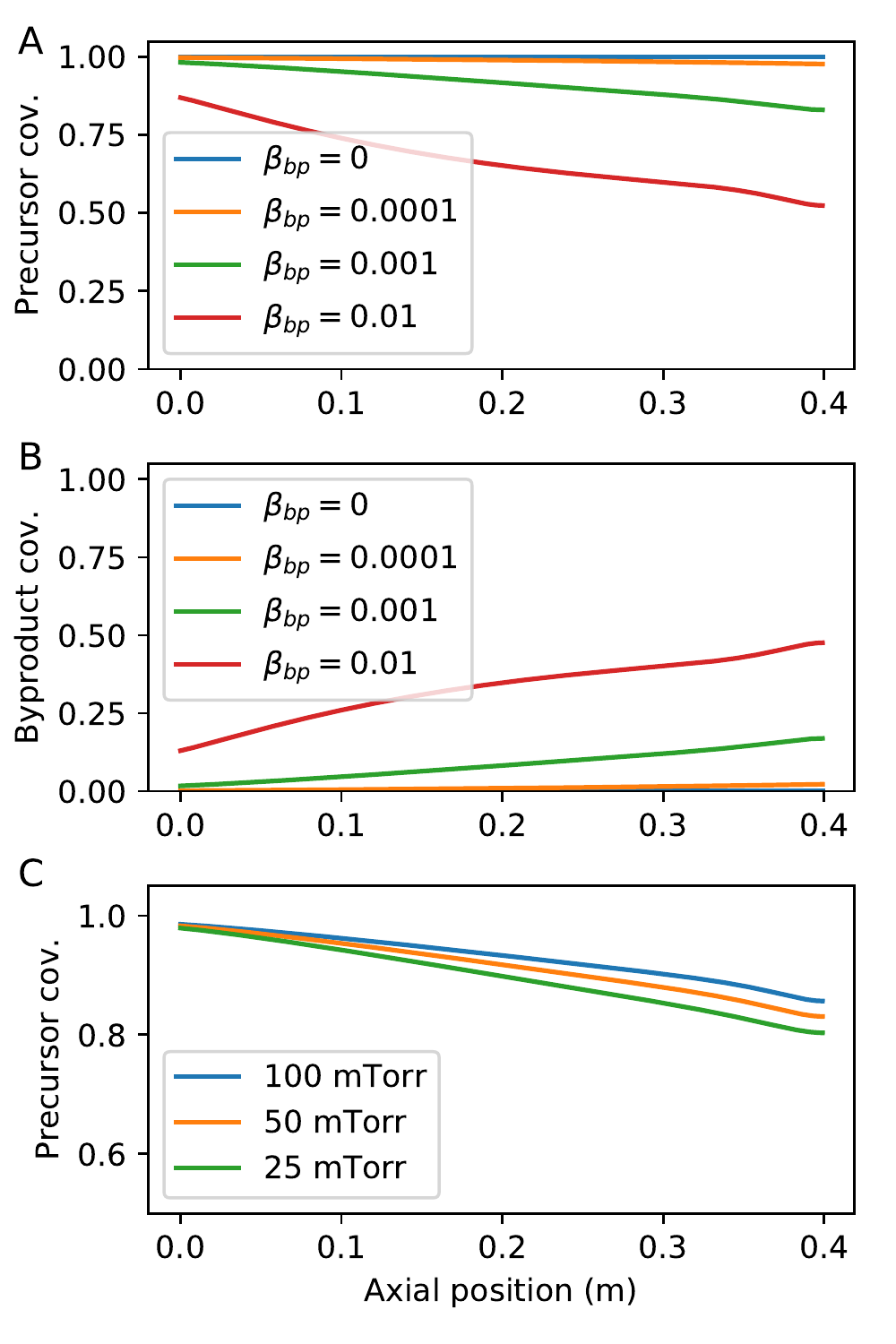}
\caption{\label{fig_tubesite}  Coverage profiles in presence of competitive adsorption by  reaction byproducts (A) Precursor coverage profile
for a process characterized by $\beta_{10} = 10^{-2}$ and $p_0=50$ mTorr. (B) Byproducts coverage profiles for the same conditions used in (A). Note
how  $\Theta + \Theta_{bp} = 1$, indicating that the process is full saturated.  (C) Impact of precursor partial pressure for a process characterized
by $\beta_{10} = 10^{-2}$ and $\beta_{bp0} = 10^{-3}$}
\end{figure}

In Figure \ref{fig_tubesite}
we show the impact that competitive adsorption has
on the saturation profiles in our tube reactor.
We consider that, on average, one ligand is released per precursor molecule.
The byproduct reactivity leads to the presence of thickness gradients even
when the process reaches saturation. These gradients
increase with the byproduct reaction probability, and
are mitigated with increasing precursor pressures. The relative diffusivities
of the precursor and byproduct molecules also play a role, controlling the
relative spread of the concentration gradients of both species as they move
downstream in the reactor.

The results in Figure \ref{fig_tubesite} have
been obtained assuming a reaction probability for the precursor of $\beta_{10}=10^{-2}$.
In Figure \ref{fig_tubesite}(A) we have explored the impact of increasing
reactivity of the reaction byproduct on the surface coverage of precursor
molecules. The dose time (0.8 s) and average precursor pressure at the
inlet during the dose (50 mTorr) lead to a complete saturation of the surface, so that,
in absence of competition, the surface coverage is one everywhere in the reactor ($\Theta=1$). As
we increase the reactivity of the surface byproduct, we observe increasing
gradients along the reactor, due to the fact that, in saturation, $\Theta = 1-\Theta_{bp}$. In Figure \ref{fig_tubesite}(B) we show the
surface coverage of the reaction byproduct $\Theta_{bp}$. The fact that the curves obtained
mirror those of the precursor shown in Fig. \ref{fig_tubesite}(A) indicate
that $\Theta + \Theta_{bp} = 1$, as expected from a fully saturated process.
In Figure \ref{fig_tubesite}(C) we show the impact of precursor pressure for
the case of $\beta_{bp0} = 10^{-3}$ under the same conditions used for
Figure \ref{fig_tubesite}(A): a higher precursor pressure tends to reduce the
impact of byproduct adsorption, as expected from a competitive adsorption process.

\begin{figure}
\includegraphics[width=2.5in]{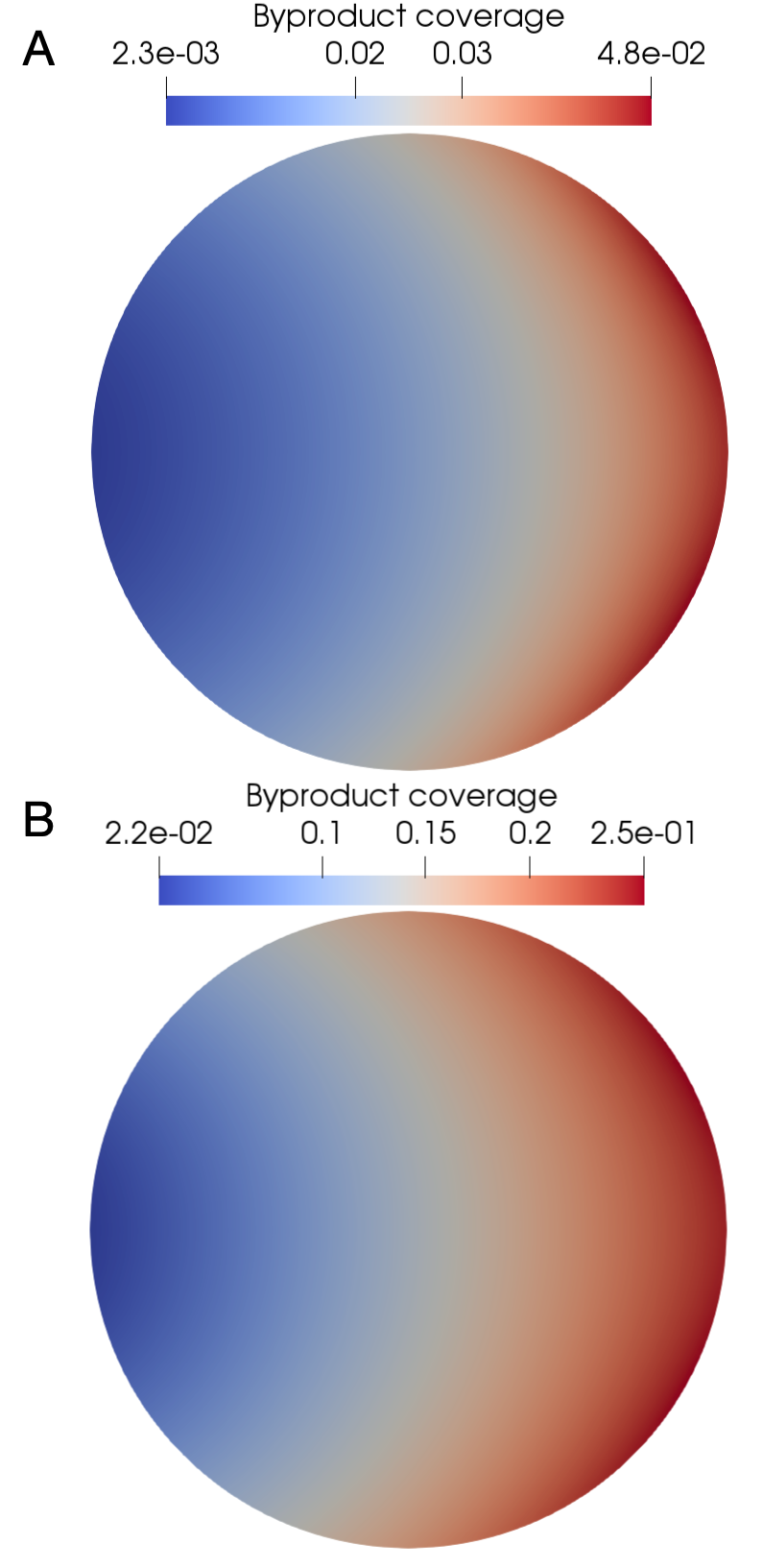}
\caption{\label{fig_byproduct} Byproduct coverage at saturation on 300mm wafers
for two
different values of byproduct reactivity: (A) $\beta_{bp} = 10^{-4}$ and (B)
$\beta_{bp} = 10^{-3}$. The precursor reactivity is $\beta_{10}=10^{-2}$.}
\end{figure}

Similar results are obtained on the 300mm wafer reactor. In
Figure \ref{fig_byproduct} we show the byproduct surface coverage at saturation for 
a precursor reactivity $\beta_{10}=10^{-2}$ and two different ligand reactivities:
$\beta_{bp} = 10^{-4}$ and $\beta_{bp} = 10^{-3}$.
Byproduct coverages increase from the upstream (left) to the downstream (right)
position in the wafer, reaching maximum values of 5\% and 25\%, respectively.

\begin{figure}
\includegraphics[width=3.5in]{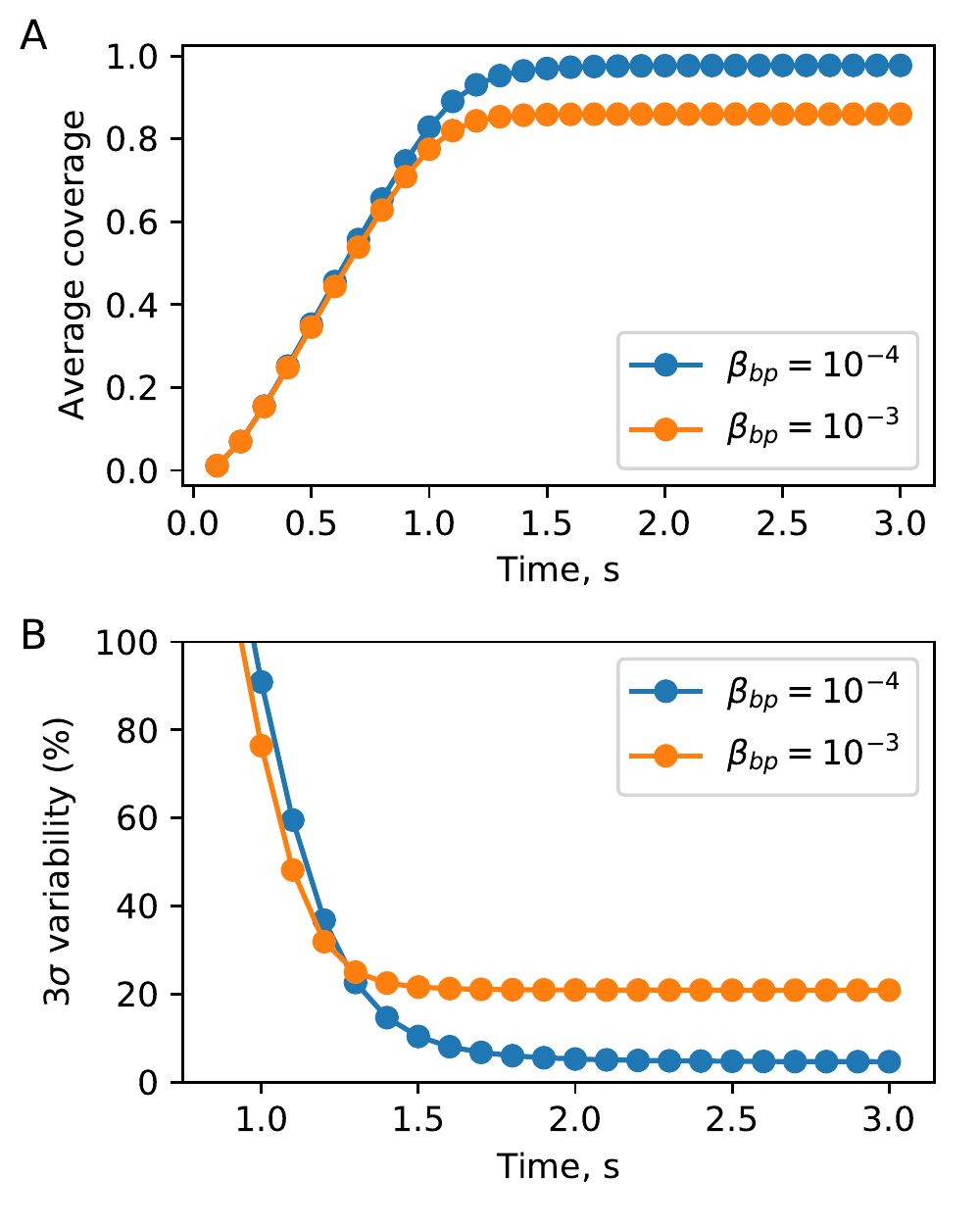}
\caption{\label{fig_byrms} Evolution of the average coverage (A) and $3\sigma$ variability in 300mm wafers for two values of byproduct reaction probability: $\beta_{bp} = 10^{-4}$ and
$\beta_{bp} = 10^{-3}$. The precursor reactivity is $\beta_{10}=10^{-2}$.}
\end{figure}

In Figure \ref{fig_byrms} we show
the evolution of precursor coverage and $3 \sigma$ variability for both
cases, showing how both processes saturate at a
point where $3\sigma$ is not zero. This provides a key differentiating feature
between a soft-saturating process and a process with competitive adsorption by reaction by products: in the former case, the $3\sigma$ 
variability is expected to decrease, albeit slowly, with increasing dose times, whereas the variability in the latter remains constant once saturation
has been reached.

\section{Discussion}

In this work we have explored self-limited processes in
two different types of reactors: a cross-flow horizontal
tube reactor and a model reactor for 300 mm wafers. For the
tube reactor case, the results obtained are in
good agreement with the prediction of an analytic model
that we previously developed under the plug-flow
approximation.\cite{YanguasGil2014}
The effect of axial diffusion is
to smooth the growth profile for undersaturated doses
with respect to the plug flow approximation. The plug
flow model also shows a reasonable agreement with
the expected QCM profiles as long as upstream consumption
and propagation delays are properly accounted for. The plug flow approximation
underestimates the initial
rise in surface coverage compared to the CFD model considering
both axial and radial diffusion, but the overall agreement
is good. 

The first order irreversible Langmuir kinetics used for the
ideal model is the simplest example of surface kinetics
exhibiting self-limited behavior. In reality, though,
many ALD and ALE processes exhibit more complex behavior.
In this work, we considered two types of
generalizations: we considered soft-saturating cases,
which we modeled taking into account fast and slow
self-limited reaction pathways, and the presence of 
site blocking by reaction byproducts. Both cases represent
examples of self-limited processes that can lead to
reactor inhomogeneities. In the first case, inhomogeneities
arise due to the timescale of the slow component requiring
unfeasibly large doses to fully saturate the surface.
The site-blocking case is intrinsically inhomogeneous even
under saturation conditions. 

The examples explored in this work are just
two of the many sources of non-ideal behavior. For instance,
it has been postulated that the effective sticking probability
of TMA can have a dependence with
precursor pressure due to the presence
of slow, reversible intermediates on the surface, something
that is also well-known from the CVD literature.\cite{Travis_TMA_2013} One
of the challenges of including increasingly complex
processes is the larger number of parameters, most
of them unknown, that are introduced in the kinetic
model:
a self-limited process that is soft-saturating
and has a slow reversible intermediate would have at
least four independent parameters per precursor. In some
cases, the surface mobility of the adsorbed species can
greatly impact the effective sticking probability, for
instance increasing the capture zone around islands
and steps in otherwise passivated surfaces.\cite{YanguasGil_markovreaction_2018}

\begin{figure}
\includegraphics[width=3in]{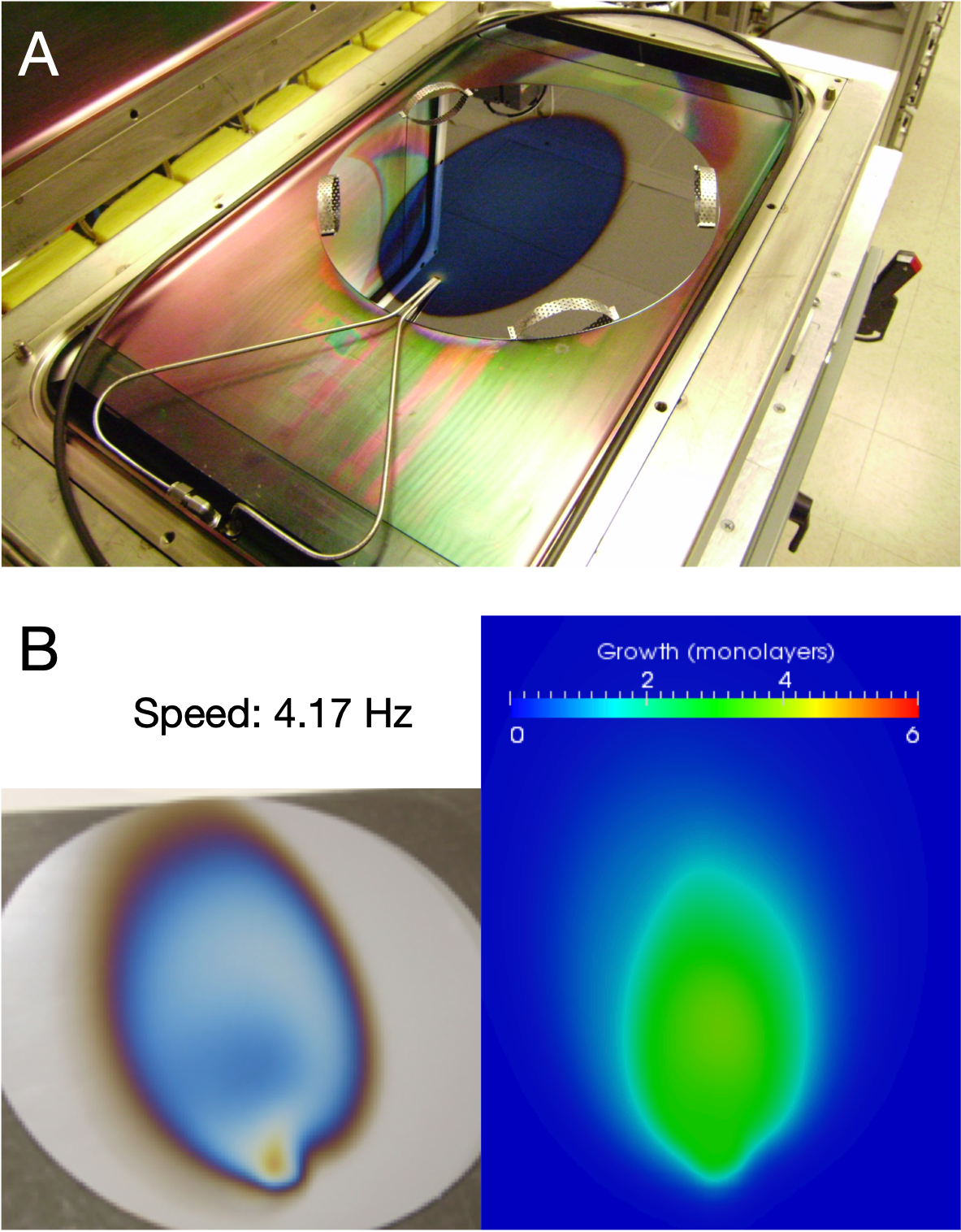}
\caption{\label{fig_injector} Comparison between simulations and
experimental results on a non-traditional experimental setup in which
sub-saturating pulses of TMA and  H$_2$O are introduced locally using a
pair of injectors. (A) Experimental setup, showing a wafer coated
at a speed of 1 cycle per second. (B) Comparison between experiments and
simulations for a speed of 4.17 cycles per second. At this rate purge times
are so short that there is an overimposed CVD component. This is captured in
the simulation results.}
\end{figure}

One of the challenges of modeling the surface kinetics
of self-limited processes is the scarcity of
kinetic data. This is also an issue for the diffusivities of different precursor molecules. Except for a few
of the most commonly used precursors and some
recent studies,\cite{Rosner,Sperling_bubbler_2019} the diffusivity
of the majority of relevant species for ALD and ALE is
not known.
One way of extracting information on surface
kinetics is through the use of saturation profiles. As
mentioned above, plug flow approximations have been
used in the past to model and extract kinetic data
from saturation profiles at a reactor scale. Likewise,
growth profiles inside high aspect ratio features can
be used to extract effective sticking probabilities
of self-limited processes. Recently, the use of macrotrenches
has greatly simplified the characterization of these
growth profiles, allowing the use of techniques such as spectroscopic ellipsometry.\cite{Karsten_2019}
The comparison between
the kinetics at a reactor scale and inside macrotrenches
can also provide information on the impact of the
surface fluxes of different species on the kinetics
of self-limited processes.

One example of the use of 
saturation profiles to extract kinetic data is shown
in Figure \ref{fig_injector}. In Fig. \ref{fig_injector}(A)
we show an experimental configuration used to explore
ALD at rates exceeding 1 cycle per second (1 Hz). Both
precursors are brought into the reactor using two long injector lines. Precursor doses are so short that it reaches 100\% precursor consumption across a 300mm wafer. 
Using kinetic data
for trimethylaluminum previously
extracted in our tube reactor and
fitted to the plug flow model, we simulated this process
using the model described in this work. The results,
shown in Fig. \ref{fig_injector}(B), are obtained after
considering three full ALD cycles. Without any fitting
parameter, the agreement between the simulation and
experiments is remarkably good, even capturing the presence of a CVD component at the center of the oval.  The main discrepancy between the simulations and the experiments 
is near the injection point: our model considers well-separated pulses, while in reality, the long injection channels cause a spread of the trimethylaluminum
and water pulses, leading to a higher growth rate near
the injectors due to precursor mixing. Still, Fig. \ref{fig_injector} provides
a good example of the potential of using reactor growth
profiles as a source for kinetic information of self-limited processes.

Also, it is important to emphasize that,
while we have focused on the case of ALD, the results
presented in this work can be directly applied to
atomic layer etching. For instance, if in Figure \ref{fig_qcm} we show
the changes in mass as a function of time and reactor position as
measured using in-situ quartz crystal microbalance. In the case
of thermal ALE, the only difference is that instead of mass gains, the
plots in Fig. \ref{fig_qcm} would represent mass losses. Likewise,
the simple 1D models explored in this work would naturally extend to the
case of thermal atomic layer etching, providing a simple way of exploring
the effective rate coefficients in ALE from etching profiles using
sub-saturating doses. 

Finally, in this work we have not considered the
effect of high surface area materials inside the reactor.
This is a technologically relevant case, as
one of the key advantages of self-limited processes is
its conformality, and yet the scale up of ALD processes to
large area substrates can sometimes be far from trivial.
In a prior work we briefly explored the impact of
a specific type of high surface area substrate on the reactive transport
of an ALD precursor.\cite{Yanguas_Gil_2014} That simple example showed the formation of a large region depleted of precursor near the center 
of the substrate for cross-flow reactors. However,
this is an area that still needs further exploration,
and will be the focus of a future work. This is also an area where
the symmetry between ALD and ALE breaks down: as the number of cycles
starts to affect the shape and surface area of the nanostructures, we expect
to see a divergence in precursor transport at reactor scale for each
case.

\section{Conclusions}

We have developed a versatile reactor scale simulation
tool capable of modeling self-limited processes such
as atomic layer deposition and atomic layer etching.
In addition to the traditional ideal self-limited
model, we have explored soft-saturating processes and
the case of competitive adsorption by reaction byproducts.
All the process, from mesh generation to the visualization
of 3D results, is based on open source tools. We have
release the simulation code as well
under a GPL v3 license and can be found at https://github.com/aldsim/aldFoam.

\begin{acknowledgments}
This research is based upon work supported by  Laboratory Directed Research and Development (LDRD) funding from Argonne National Laboratory, provided
by the Director, Office of Science, of the U.S. Department of Energy under Contract No. DE-AC02-06CH11357.
We gratefully acknowledge the computing resources provided on Blues, a high-performance computing cluster operated by the Laboratory Computing Resource Center at Argonne National Laboratory. We would also like to acknowledge
Shashikant Aithal for his support and useful suggestions for
testing our simulations in Blues.

\end{acknowledgments}

\bibliography{aldsimulations}

\end{document}